\documentclass[twocolumn]{aastex631} 

\usepackage{newtxtext,newtxmath,comment} 
\usepackage[normalem]{ulem}

\usepackage[T1]{fontenc}
\usepackage{ae,aecompl}
\usepackage{xcolor}
\usepackage{graphicx}
\usepackage{amsmath}
\usepackage{enumitem}
\usepackage{url}
\usepackage{tasks}
\usepackage{xspace}
\usepackage{makecell}

\def\mdot{\mathrm{M}_\odot}
\def\mtot{M\mathrm{_{TOT}}}

\def\mstar{M_*}
\def\ew{$\widebar{e}_w$}
\def\kw{$\kappa_w$}
\def\om{$\Omega_\mathrm{m}$}
\def\s8{$\sigma_8$}
\def\agn{$\epsilon_{f,\mathrm{high}}$}

\defcitealias{DREAMS}{Rose25}

\makeatletter
\let\frontmatter@title@above=\relax
\makeatother

\begin{document}

\title[The DREAMS Project]{The DREAMS Project:\\
A New Suite of 1,024 Simulations to Contextualize the Milky Way and Assess Physics Uncertainties}

\correspondingauthor{Jonah C. Rose} \\
\email{jr8952@princeton.edu}

 \author[0000-0002-2628-0237]{Jonah C. Rose}
\affiliation{Department of Physics, Princeton University, Princeton, NJ 08544, USA}
\affiliation{Center for Computational Astrophysics, Flatiron Institute, 162 5th Avenue, New York, NY 10010, USA}

\author[0000-0002-8495-8659]{Mariangela Lisanti}
\affiliation{Department of Physics, Princeton University, Princeton, NJ 08544, USA}
\affiliation{Center for Computational Astrophysics, Flatiron Institute, 162 5th Avenue, New York, NY 10010, USA}

\author[0000-0002-5653-0786]{Paul Torrey}
\affiliation{Department of Astronomy, University of Virginia, 530 McCormick Road, Charlottesville, VA 22904}
\affiliation{Virginia Institute for Theoretical Astronomy, University of Virginia, Charlottesville, VA 22904, USA}
\affiliation{The NSF-Simons AI Institute for Cosmic Origins, USA}

\author[0000-0002-4816-0455]{Francisco Villaescusa-Navarro}
\affiliation{Center for Computational Astrophysics, Flatiron Institute, 162 5th Avenue, New York, NY 10010, USA}

\author[0000-0002-8111-9884]{Alex M. Garcia}
\affiliation{Department of Astronomy, University of Virginia, 530 McCormick Road, Charlottesville, VA 22904}
\affiliation{Virginia Institute for Theoretical Astronomy, University of Virginia, Charlottesville, VA 22904, USA}
\affiliation{The NSF-Simons AI Institute for Cosmic Origins, USA}

\author[0000-0003-0777-4618]{Arya~Farahi}
\affiliation{Departments of Statistics and Data Sciences, University of Texas at Austin, Austin, TX 78757, USA}
\affiliation{The NSF-Simons AI Institute for Cosmic Origins, USA}

\author[0000-0001-5522-5029]{Carrie~Filion}
\affiliation{Center for Computational Astrophysics, Flatiron Institute, 162 5th Avenue, New York, NY 10010, USA}

\author[0000-0002-0372-3736]{Alyson M. Brooks}
\affiliation{Department of Physics \& Astronomy, Rutgers, the State University of New Jersey, Piscataway, NJ 08854, USA}

\author[0000-0002-3204-1742]{Nitya Kallivayalil} 
\affiliation{Department of Astronomy, University of Virginia, 530 McCormick Road, Charlottesville, VA 22904}
\affiliation{The NSF-Simons AI Institute for Cosmic Origins, USA}

\author[0000-0003-4004-2451]{Kassidy E. Kollmann}
\affiliation{Department of Physics, Princeton University, Princeton, NJ 08544, USA}

\author[0009-0000-8180-9044]{Ethan Lilie} 
\affiliation{Department of Physics, Princeton University, Princeton, NJ 08544, USA}

\author[0000-0001-7168-8517]{Bonny Y. Wang} 
\affiliation{Department of Astronomy and Astrophysics, University of Chicago, Chicago, IL 60637, USA}

\author[0000-0001-7831-4892]{Akaxia Cruz}
\affiliation{Center for Computational Astrophysics, Flatiron Institute, 162 5th Avenue, New York, NY 10010, USA}
\affiliation{Department of Physics, Princeton University, Princeton, NJ 08544, USA}

\author[0000-0002-7638-7454]{Sandip Roy} 
\affiliation{Department of Astronomy \& Astrophysics, University of California, San Diego, La Jolla, CA 92093, USA}

\author[0000-0002-6021-8760]{Andrew B. Pace}
\affiliation{Department of Astronomy, University of Virginia, 
530 McCormick Road, 
Charlottesville, VA 22904}

\author[0009-0002-1233-2013]{Niusha Ahvazi} 
\affiliation{Department of Astronomy, University of Virginia, 530 McCormick Road, Charlottesville, VA 22904}
\affiliation{Virginia Institute for Theoretical Astronomy, University of Virginia, Charlottesville, VA 22904, USA}
\affiliation{The NSF-Simons AI Institute for Cosmic Origins, USA}

\author[0000-0002-7968-2088]{Stephanie O'Neil} 
\affiliation{Department of Physics \& Astronomy, University of Pennsylvania, Philadelphia, PA 19104, USA}
\affiliation{Department of Physics, Princeton University, Princeton, NJ 08544, USA}

\author[0000-0002-3400-6991]{Cian Roche}
\affiliation{Department of Physics and Kavli Institute for Astrophysics and Space Research, Massachusetts Institute of Technology, Cambridge, MA 02139, USA}

\author[0000-0002-6196-823X]{Xuejian Shen} 
\affiliation{Department of Physics and Kavli Institute for Astrophysics and Space Research, Massachusetts Institute of Technology, Cambridge, MA 02139, USA}


\author[0000-0001-8593-7692]{Mark Vogelsberger}
\affiliation{Department of Physics $\&$ Kavli Institute for Astrophysics and Space Research, Massachusetts Institute of Technology, Cambridge, MA 02139, USA}

\begin{abstract}
We introduce a new suite of 1,024 cosmological and hydrodynamical zoom-in simulations of Milky Way-mass halos, run with Cold Dark Matter, as part of the DREAMS Project.  Each simulation in the suite has a unique set of initial conditions and combination of cosmological and astrophysical parameters.  The suite is designed to quantify theoretical uncertainties from halo-to-halo variance, as well as stellar and black hole feedback. 
We develop a novel weighting scheme that prioritizes regions of the input parameter space, yielding galaxies consistent with the observed present-day stellar mass--halo mass relation. 
The resulting galaxy population exhibits a wide diversity in structural properties that encompasses those of the actual Milky Way, providing a powerful statistical sample for galactic archaeology.
To demonstrate the suite's scientific utility, we investigate the connection between a galaxy's merger history, focusing on Gaia-Sausage-Enceladus~(GSE) analogs, and its present-day properties.
We find that galaxies with a GSE analog have lower star formation rates, more compact disks, and more spherical stellar halos.
Crucially, significant halo-to-halo scatter remains, demonstrating that matching more than the most significant events in the Milky Way's past is necessary to recover its present-day properties.
Our results highlight the necessity for large statistical samples to disentangle the stochastic nature of galaxy formation and robustly model the Milky Way's unique history.
\end{abstract}


\section{Introduction}
\label{sec:intro}

The last decade has marked a golden age for galactic archaeology.
Surveys like \textit{Gaia}~\citep{2016Gaia, 2023Gaia}, the Dark Energy Survey~\citep[DES;][]{2005DES,2022DES}, and the Apache Point Observatory Galactic Evolution Experiment~\citep[APOGEE;][]{2017APOGEE, 2022APOGEE} have provided an unprecedented view of the Milky Way~(MW) and its components. Taken together, these data have accelerated the mapping of the chemodynamical structure of the Galaxy's stellar disk, bulge, and halo~\citep[e.g.,][]{2025NewAR.10001721H}.  They have also led to the discovery of new satellite galaxies, as well as their tidal debris~\citep[e.g.,][]{2019ARA&A..57..375S, 2020ARA&A..58..205H}.  This data-rich landscape will only continue to improve with upcoming results from the Vera~C. Rubin Observatory~\citep{2019LSST}, the Nancy Grace Roman Space Telescope~\citep{2015Spergel}, and others. 

As we map our Galaxy to increasing precision, a natural question that arises regards how typical its properties are compared to other systems of similar mass.  It is becoming possible to place the MW in a cosmological context by comparing it to millions of external galaxies found in surveys like the Sloan Digital Sky Survey~\citep[SDSS;][]{2023SDSS}, the Galaxy and Mass Assembly Survey~\citep[GAMA;][]{2011GAMA, 2015GAMA}, and the Hyper Suprime Cam Survey~\citep[HSC;][]{2018HSC,2022HSC}. While these extragalactic surveys have not provided the same detail for other galaxies as our own, they have measured the stellar mass~\citep{2003Kauffmann,2011Taylor,2013Ilbert,2022Driver}, metallicity~\citep{2004Tremonti,2008Kewley,2010Mannucci,2013Lara}, morphology~\citep{2011Lintott,2012Kelvin,2018Dominguez}, and more.
Comparing results across datasets is challenging due to varying methodologies, such as different observed tracers or sample selection criteria~\citep[e.g.,][]{2014Speagle}, and different sources of uncertainty, such as dust modeling or assumed mass-to-light ratios~\citep[e.g.,][]{2013Conroy}, but the comparison is crucial for providing context for the Galaxy in which we live.

Through these comparisons, some features of our Galaxy appear to be rare when compared to a broader population of MW analogs.
For instance, our stellar disk has been found to be both more massive and more compact than those found in galaxies of comparable mass~\citep{2016Bland, 2016Licquia}.
The supermassive black hole~(SMBH) at the Galactic Center is also less massive than that of other MW-mass galaxies~\citep[for a review, see][]{2013Kormendy}, although it is more similar to other late-type galaxies~\citep{2020Greene}.
Additionally, features outside the central Galaxy, such as the presence of a close and massive satellite, the Large Magellanic Cloud~(LMC), and a massive neighbor, Andromeda~(M31), are relatively uncommon~\citep{2010Boylan-Kolchin,2011Guo}.

Properly contextualizing the MW requires robust theoretical models with large galaxy samples to compare to data.
Large-volume simulations, such as IllustrisTNG~\citep{2018Pillepicha, 2017Weinberger, 2018Weinberger} and EAGLE~\citep{2015Schaye,2015Crain,2016McAlpine}, provide statistical samples of galaxies to understand the rarity of particular features and how they correlate with other galactic properties.
While providing additional information on the environment, their statistical power comes at the cost of resolution, which limits their ability to model the internal structure and formation history of individual galaxies.
On the other hand, simulations that focus on individual galaxies, so-called `zoom-in' simulations, can achieve higher resolutions needed to model disks, bulges, and satellite properties~\citep[for a review, see][]{2020Vogelsberger}.  
However, a large number of zoom-in simulations are required to disentangle robust physical trends from the stochasticity of halo-to-halo variance, as each zoom-in simulation generally models one galaxy, quickly becoming computationally challenging for large samples.  
This is further complicated by specific aspects of the MW's merger history, such as the presence of the LMC, which directly affect properties such as satellite number counts~\citep{2020Nadler,2021DSouza,2024Buch} and the structure of its stellar halo~\citep{2019Garavito,2021Conroy,2024Arora}.




In addition to the inherent uncertainty caused by halo-to-halo variance, galaxy simulations are subject to the uncertainty in the sub-grid models that are used to account for unresolved physics.
Many of the large-scale state-of-the-art simulations tune their models to match observed scaling relations like the stellar mass--halo mass relation~\citep[SMHM;][]{2014Torrey,2015Schaye,2018Pillepicha,2021Lovell}.
Yet, many of these tuned parameters are degenerate for a given scaling relation~\citep{SB28} and can introduce unquantified theoretical uncertainty.

New approaches to large-scale simulations, such as the Cosmology and Astrophysics with MachinE Learning Simulations~\citep[CAMELS;][]{CAMELS,CMD} and the DaRk MattEr and Astrophysics with Machine learning and Simulations Project~\citep[DREAMS;][]{DREAMS}, address this by no longer relying on a single set of tuned parameters.
The DREAMS Project in particular generates thousands of simulations, varying over both halo initial conditions and simulation parameters to incorporate more realistic theoretical uncertainties, accounting not only for halo-to-halo variance, but also incomplete understanding of the underlying galaxy-formation physics.

In this paper, we introduce a new suite of 1,024 cosmological and hydrodynamical simulations of MW-mass systems as part of the DREAMS~Project\footnote{\url{https://dreams-project.readthedocs.io}\\ \url{https://www.dreams-project.org}}~\citep{DREAMS}.  Unlike the Warm Dark Matter~(WDM) suites presented in \cite{DREAMS}, these simulations assume a Lambda Cold Dark Matter~($\Lambda$CDM) cosmology.  This paper presents the simulation suite (Section~\ref{sec:methods}), which varies key parameters in the TNG galaxy-formation model, and introduces a novel weighting scheme (Section~\ref{sec:weights}) to constrain the multi-dimensional parameter space by comparing to the observed SMHM.  This then allows us to perform a detailed examination of the central MW-mass galaxies in the suite, comparing them to properties of our own Galaxy (Section~\ref{sec:mw_props} and Section~\ref{sec:gse}).  We conclude in Section~\ref{sec:conclusion}.  
Finally, Appendices~\ref{app:nehod}~and~\ref{app:props} present a validation of our emulators, supplementary figures, and more details on disentangling the structural imprints of the Gaia-Sausage-Enceladus~(GSE) merger from the stochastic variance driven by baryonic feedback.

\section{Simulations}
\label{sec:methods}

\begin{table*}
    \centering
    \begin{tabular}{llllllllll}
        \Xhline{3\arrayrulewidth} 
        Dark Matter        & \# of & Baryon       &  \om          & \s8           & \ew        & \kw        &  \agn\,      & Target Mass       & Notes \\
        Prescription       & Sims  & Prescription &               &               &            &            &             & [$10^{12}~\mdot$] &       \\
        \hline
        CDM                & 1024  & N-body       & [0.274, 0.354] & [0.780, 0.888] &    -       &     -      &       -     & [0.50, 2.00]    & This paper  \\
                           & 1024  & TNG          & [0.274, 0.354] & [0.780, 0.888] & [0.9, 14.4] & [3.7, 14.8] & [0.025, 0.4] & [0.50, 2.00]    & This paper  \\
        \hline
        WDM                & 1024  & N-body       & 0.302         & 0.839         &    -       &     -      &      -      & [1.00, 1.85]    &~\citetalias{DREAMS} \\
                           & 1024  & TNG          & 0.302         & 0.839         & [0.9, 14.4] & [3.7, 14.8] & [0.025, 0.4] & [1.00, 1.85]    &~\citetalias{DREAMS} \\
        \Xhline{3\arrayrulewidth}      
    \end{tabular}
    \caption{Summary of the DREAMS MW-mass simulation suites. 
    The table outlines the number of simulations, the DM and baryonic physics prescriptions used, the ranges of varied parameters, the target halo mass range, and the corresponding reference papers. 
    The varied parameters are the total matter density, $\Omega_{\rm m}$, the amplitude of matter fluctuations, $\sigma_8$, the specific energy of SN feedback, $\widebar{e}_w$, the SN wind velocity, $\kappa_w$, and the coupling of AGN feedback to the surrounding gas, $\epsilon_{f,{\rm high}}$. The WDM suites are from~\cite{DREAMS}, shown as~\citetalias{DREAMS}. 
    The hydrodynamical CDM suite introduced in this work has a DM particle mass of $1.8\times(\Omega_{\rm m}/0.314)\times10^{6}~\mdot$, a baryon particle mass of $2.8\times10^5~\mdot$, and a gravitational softening of 0.441~kpc. Each of the 1,024 simulations starts from a unique initial condition. Each initial condition in the hydrodynamic suite has an N-body counterpart with the same resolution, but where the baryons are treated as collisionless particles. The hydrodynamical simulations cost 8.7 million CPU hours, and the corresponding N-body suite cost 1.9~million CPU hours. This paper does not make explicit use of the CDM N-body simulations, so all references to the DREAMS CDM suite refer specifically to the hydrodynamical one.
    }
    \label{tab:sims}
\end{table*}

We introduce a new suite of DREAMS zoom-in hydrodynamical simulations, which target MW-mass galaxies in a $\Lambda$CDM cosmology. It comprises 1,024 unique cosmological simulations that explore theoretical modeling uncertainties by varying parameters related to the cosmology and TNG sub-grid physics~\citep{2018Pillepicha, Weinberger2020}.
In addition, we include a complementary suite of 1,024 N-body simulations that start from the same initial conditions as the hydrodynamic suite but model the baryons as collisionless.
The suites also explore the effects of halo-to-halo variance as each simulation is generated from a unique initial density field.  This section reviews the relevant details of the CDM suites, which are summarized in Table~\ref{tab:sims}, and also highlights the differences in methodology from the previously published DREAMS MW-mass WDM suites~\citep{DREAMS}.

\subsection{Simulation Setup and Parameters}
\label{sec:sims}

All simulations are generated with the moving mesh code \textsc{Arepo}~\citep{Springel2010, Springel2019, Weinberger2020}.
We use the \textsc{subfind} algorithm~\citep{2001Springel} to locate galaxies in the simulations.
The simulation procedure follows that described in Appendix~A of~\cite{DREAMS} with some minor adjustments outlined in Section~\ref{sec:updates}.
A summary of the zoom-in procedure is provided here for reference.

First, we generate initial conditions for an N-body simulation with a side length of 145~cMpc at $z=127$ using \textsc{music}~\citep{2011Hahn}.
The simulation is evolved to $z=0$ with \textsc{Arepo} at a particle mass resolution of $7.2\times10^9~\mdot$.
From this parent simulation, a random halo with a total mass between $(0.50$--$2.00)\times10^{12}~\mdot$ is regenerated in an N-body simulation with a particle mass resolution of $1.5\times10^7~\mdot$.
Here, total mass refers to M$_{\rm 200}$, the mass enclosed in a sphere whose mean density is 200$\times$ the critical density of the Universe.
If the re-simulated halo does not fall within the target mass range or has another halo more massive than $1.0\times10^{12}~\mdot$ within 1~Mpc, then a different halo is chosen from the uniform box.
This isolation criterion is imposed to make the simulations computationally feasible to complete.
If the target halo is within the mass and isolation limits, then it is resimulated with TNG physics with DM resolution of $1.8\times(\Omega_{\rm m}/0.314)\times10^{6}~\mdot$, a baryon resolution of $2.8\times10^5~\mdot$, and a gravitational softening of 0.441~kpc.

These simulations have $4\times$ worse mass resolution than the TNG50 box~\citep{2019Pillepich}, the highest-resolution simulation of the IllustrisTNG suite.
However, the DREAMS suites have $5\times$ the number of MW-mass galaxies and are chosen from box volumes that are $24\times$ larger, allowing for a greater range of environments and formation histories.
The isolation criterion limits DREAMS halos to lower-density environments; \cite{DREAMS} found that approximately 12\% of potential target halos fail this criterion.  The relative isolation of the MW-mass galaxies in the current DREAMS suites should be taken into account when comparing to our Galaxy; in particular, M31 is not modeled, nor do we require that the simulated MWs have an LMC or GSE.  

The TNG model includes key physical processes, such as metal-line cooling, star formation, and feedback from supernovae~(SNe) and active galactic nuclei~(AGN)~\citep{2018Pillepicha}.
For a detailed discussion of the TNG model in the DREAMS simulations, see Section~2.1 in~\cite{DREAMS}.
To explore theoretical uncertainties, we vary three astrophysical (two SN and one AGN) and two cosmological parameters.

The SN parameters are related through the mass-loading factor, $\eta_w$, which measures the ratio of gas that is removed from the galaxy by stellar winds and star formation.  It is defined by
\begin{equation}
    \label{eq:mass_loading}
    \eta_w = \frac{2}{v_w^2} e_w \left(1 - \tau_w \right) \, ,
\end{equation}
where $\tau_w$ is the fraction of energy released thermally (fixed to the fiducial TNG value of $0.1$). In Equation~\ref{eq:mass_loading}, 
$e_w$ is the specific energy available to generate winds, given by
\begin{align}
    e_w =&~ \bar{e}_w\cdot f(Z)\cdot N_{\rm SN}\,\left[10^{51}\frac{\rm erg}{{\rm M}_\odot}\right] \,,
\end{align}
where the dimensionless parameter $\bar{e}_w$ characterizes the energy per
SN, the function $f(Z)$ reduces that energy in a metallicity-dependent manner, and $N_{\rm SN}$ is the number of Type~II SNe. The velocity of the stellar winds, $v_w$, also appears in Equation~\ref{eq:mass_loading} and is given by 
\begin{equation}
    v_w = \mathrm{max} \left[\kappa_w \sigma_{\scriptscriptstyle{\rm DM}} \left(\frac{H_0}{H(z)} \right)^{1/3}, \, v_{w,\mathrm{min}} \right] \, ,
\end{equation}
where $\kappa_w$ is a dimensionless normalization factor, $\sigma_{\scriptscriptstyle{\rm DM}}$ is the local velocity dispersion associated with the 64-closest DM particles, $H(z)$ is the Hubble parameter at redshift $z$ (designated as $H_0$ for present day), and $v_{w,\mathrm{min}} = 350$~km/s is the minimum wind speed. 

The two SN parameters that we vary, \kw\,and $\widebar{e}_w$, are unitless multiplicative factors that apply to the wind velocity and specific energy, respectively.  
We sample \kw\,uniformly over the logarithmic range $\kappa_w \in [3.7, 14.8]$, corresponding to a factor of two variation about the TNG fiducial value of 7.4.  
We sample \ew\,uniformly over the logarithmic range $\widebar{e}_w\in [0.9, 14.4]$, corresponding to a factor of four variation about the TNG fiducial value of 3.6.

The AGN parameter that we vary, $\epsilon_{f,{\rm high}}$, controls the fraction of thermal energy that is transferred from the AGN to nearby gas.
Black holes~(BHs) are seeded with a mass of $1.2\times10^6~\mdot$ once a halo reaches a mass of $7.2\times10^{10}~\mdot$.
We note that the large seed mass already comprises $\sim$$25\%$ of the mass of Sagittarius~A*~\citep{2016Bland}, creating a high mass floor when comparing to observed values.
There are two feedback modes in the TNG model: a low-accretion state corresponding to a system where hot gas isotropically accretes onto the BH~\citep{1976ApJ...204..187S,1977ApJ...214..840I} and a high-accretion state with cooler gas accreting through a disk~\citep{1973A&A....24..337S}.
The high-accretion state dominates for galaxies with lower stellar masses, $\mstar<10^{10.5}~\mdot$, making it the more relevant for MW-mass halos.
In this mode, the overall energy released from accretion onto the AGN, $\dot{E}_{\text{AGN}}$, is given by
\begin{equation}
    \dot{E}_{\text{AGN}} = \epsilon_{f,\mathrm{high}} \epsilon_r \dot{M}_{\scriptscriptstyle{\rm BH}} c^2 \, ,
\end{equation}
where $\epsilon_r$ is the radiative efficiency,  $\dot{M}_{\scriptscriptstyle{\rm BH}}$ is the accretion rate of gas onto the AGN, and $c$ is the speed of light.
We sample \agn\,uniformly over the logarithmic range $\epsilon_{f,\mathrm{high}}\in [0.025, 0.4]$, corresponding to a factor of four variation about the TNG fiducial value of 0.1.

The TNG model contains many, $\mathcal{O}$(100), sub-grid physics and/or numerical parameters that impact its behavior.
These parameters are heterogeneous in nature;  some correspond to physical quantities for which we have clear physical expectations (e.g., the SN energy budget is expected to be $\sim$$10^{51}~\mathrm{erg}$ per core-collapse SN) while others are not easily linked to clear physical constraints (e.g., the effective equation-of-state parameter for the ~\citealt{2003Springel} star-formation model where the stiffness is selected to ensure numerical stability and favorable overall model performance).  
The default TNG model, and previously the Illustris model, was selected by comparing performance against a series of observed galaxy properties, such as galaxy stellar mass functions, cosmic star formation rate densities, and mass-size relations---see~\citet{2013Vogelsberger, 2014Torrey, 2018Pillepicha}. However, the default TNG model parameter choices are not necessarily a uniquely good setup (i.e., there may be other sets of parameter selections that do equally well at matching observed galaxy properties) and, indeed, may not even reflect the globally optimal model parameter choices.  Because the TNG model was tuned by comparing a modest number of variations against one another, there may very well be ``better" model setups that were not identified.  Put another way: there is some uncertainty associated with predictions of the TNG model owing to the fact that it contains both physically motivated and algorithmically motivated parameters whose true/best values are not uniquely constrained by data, allowing multiple plausible parameter/model configurations to produce comparably realistic galaxies.
Fully capturing the variability in the behavior of the TNG model would require varying a large number of parameters, as done in, e.g.,~\citet{SB28}.  In this work, we focus on only three astrophysical parameters ($e_w$, $\kappa_w$, and $\epsilon_{f,\mathrm{high}}$) as they directly impact the efficiencies of stellar and AGN feedback within the TNG framework.  While other model ingredients (e.g., the IMF, equation-of-state stiffness parameter, etc.) will also influence these feedback channels, these three parameters drive significant variability in the stellar and AGN feedback properties.  
The ranges that we choose for these parameters are taken to vary about the default values from the original TNG model~\citep{2018Pillepicha} by factors of a few, consistent with the parameter variations used in the CAMELS simulations~\citep{CAMELS, CMD}.


Along with variations to the astrophysics, we also vary two cosmological parameters: the total matter density, \om, and the amplitude of matter fluctuations, $\sigma_8$.  We sample each uniformly in the range $\Omega_{\rm m} \in [0.274, 0.354]$ and $\sigma_8\in[0.780, 0.888]$, respectively.\footnote{For reference, TNG sets $\Omega_{\rm m} =0.310$ and $\sigma_8 = 0.816$.} These ranges correspond to roughly  3$\sigma$ about the 2016 Planck results~\citep{2016Planck} and 5$\sigma$ about the 2018 Planck results~\citep{2020Planck}.
The ranges are intentionally kept wide for possible machine-learning applications.  
For example, when computing posterior probabilities in simulation-based inference~(SBI), over-constrained priors may limit the ability to infer physical results, as they can dominate the posteriors.
All other cosmological parameters, $h = 0.691$ and $\Omega_{\rm b} = 0.046$, are fixed for each simulation and are consistent with Planck 2016 values~\citep{2016Planck} and the DREAMS WDM suite~\citep{DREAMS}.


In summary, each simulation in this suite contains a unique initial density field, as well as a unique combination of five astrophysical and cosmological parameters that are varied according to a Sobol sequence~\citep{sobol}.  We simulate 1,024 simulations in each suite as this number produces precise results when training machine learning simulations in CAMELS~\citep{CAMELS}.
Appendix~\ref{app:nehod} demonstrates how the machine learning results presented in this work depend on the number of simulations, see Figures~\ref{fig:sim_count}~and~\ref{fig:sim_count_visual}. For emulating the SMHM relation, the results converge to the results of the full DREAMS suite with $\sim$200 simulations.
Additional simulation number convergence tests can be found in Rose et al.~(in prep).

As indicated in Table~\ref{tab:sims}, each hydrodynamic simulation has a corresponding N-body simulation that starts from the same initial conditions, runs at the same resolution, and has the same cosmological parameters. 
This suite, which only includes variations over the cosmological parameters \om\, and $\sigma_8$, can be used to explore how the inclusion of baryons affects each halo.  This paper does not make explicit use of the N-body simulations, so all references to the DREAMS CDM suite refer specifically to the hydrodynamical one.

\subsection{Updates to the DREAMS Framework}
\label{sec:updates}

We have implemented three key updates to the simulation framework presented in~\cite{DREAMS}.

First, the distribution of final MW-mass galaxies is more uniform over the intended mass range.
In the original WDM suite, MW-mass halos were selected from $(1.58\text{--}1.61)\times10^{12}~\mdot$ at a low resolution.
This produced a mass distribution in the final high-resolution suite with an average of $\left(1.3\pm0.12\right)\times10^{12}~\mdot$.
In this work, we update the selection method to choose halos between the masses of $\left(0.5\text{--}2.0\right)\times10^{12}~\mdot$ and ensure that the halo remains within this range through the intermediate N-body zoom-in stage.
The updated selection method produces a more uniform distribution over the current range of uncertainty on the MW Galaxy's total mass~\citep{1999Wilkinson, 2010Watkins, 2014Kafle, 2018Sohn, 2019Eadie, 2022Wang}.

Second, the method used to define the high-resolution Lagrangian region for the zoom-in simulations is updated to reduce computational time.
In the original procedure, when the Lagrangian region was created from the particles within $5R_\mathrm{200}$ at $z=0$ and produced high contamination or computational time, the region was modified by removing distant particles and padding the central region with additional particles.\footnote{$R_{\rm 200}$ corresponds to the radius within which the average density of a halo is $200$ times the critical density,}
For a full description of the procedure, see Appendix~A in~\cite{DREAMS}.
Now, the padding method is applied to every Lagrangian region that is created from all the particles within $5R_{\rm 200}$ to reduce the size of the high-resolution region.
Applying this method to all simulations results in an overall reduction in size for the largest Lagrangian regions but does not alter smaller regions.
Although uncommon, this method can result in higher low-resolution particle contamination within the target MW-mass halo, but we maintain a 2\% contamination limit by resimulating any halos that pass this threshold.
We define the contamination as the ratio of all low-resolution DM mass within $R_{\rm 200}$ over the total mass of the galaxy within $R_{\rm 200}$.

Third, the range over which each cosmological parameter is varied is reduced.
As originally presented in~\cite{DREAMS}, the WDM uniform-box suite varied $\Omega_\mathrm{m}$ and $\sigma_\mathrm{8}$ over the ranges $\Omega_\mathrm{m}\in[0.1, 0.5]$ and $\sigma_\mathrm{8}\in[0.6, 1.0]$.
We implement this change because the focus of this suite is less on constraining cosmology, but analyses may still be affected by cosmological uncertainties.
This change focuses the simulations on the parameter space of highest observational interest, providing sufficient variation to account for current cosmological tensions without exploring deeply disfavored regions of parameter space.

\section{Galaxy Weighting Procedure} 
\label{sec:weights}

A key advantage of the DREAMS Project is the ability to quantify theoretical uncertainties by exploring how variations in cosmological and astrophysical parameters affect simulation results.
Although the prior range for each parameter is chosen to align with current observational uncertainties,
certain regions of the five-dimensional parameter space likely produce galaxies inconsistent with observations. 
This section describes a procedure that compares the simulation outputs to the observed stellar mass--halo mass~(SMHM) relation for MW-mass galaxies, down-weighting regions of parameter space that produce unrealistic galaxies.  This weighting scheme can be used to identify the regions of parameter space that are most consistent with observations, the SMHM relation in this case, but also other scaling relations when generalized. Section~\ref{sec:emulator} introduces the procedure used to emulate the DREAMS dataset, which is detailed in Appendix~\ref{app:nehod}. Section~\ref{sec:weight_method} then describes the implementation of the weighting procedure, and Section~\ref{sec:unweighted} demonstrates the impact of the procedure using a recent data-driven SMHM relation from the literature.
Appendix~\ref{app:nehod} provides additional details and validation tests for the emulator used in this Section.

\subsection{Emulation and Data Generation}
\label{sec:emulator}

To efficiently explore the five-dimensional parameter space, we generate a large sample of MW-mass galaxies by emulating the DREAMS dataset.
This is achieved with a conditional normalizing flow to emulate the central galaxy properties, following an architecture similar to that used in~\cite{NeHOD}.
In general, the emulator can be conditioned on the five simulation parameters described in Section~\ref{sec:sims} ($\Omega_{\rm m}$, $\sigma_8$, $\widebar{e}_w$, $\kappa_w$, $\epsilon_{f,{\rm high}}$) and trained to predict key properties of the central galaxy, stellar mass and total halo mass, that are required for the weighting scheme calculation.

Normalizing flows are a class of generative models that learn a complex multi-dimensional data distribution by transforming a base distribution (typically Gaussian) through a series of invertible and differentiable functions~\citep[for a review, see][]{2021Papamakarios}.
The model is trained by directly maximizing the log likelihood without adversarial training required by other generative models, like generative adversarial networks~\citep{2014Goodfellow}.
For details on the emulator architecture, hyperparameter tuning, training, and validation, see Appendix~\ref{app:nehod}.

To incorporate epistemic uncertainty, the uncertainty arising from specific model training and hyperparameter choices, we construct an emulator ensemble~\citep[e.g.,][]{2017Lakshminarayanan}.
This technique combines the output from multiple trained models to make more robust predictions.
Based on our hyperparameter search, we select the ten individual models that achieve the lowest validation loss, each of which is then used to generate 10\% of the final dataset.
This approach averages the small variations between the best-performing models to provide a more stable and reliable set of predictions.
After training, we use the emulator ensemble to create a large dataset of generated galaxies, see Section~\ref{sec:weight_method}.

\subsection{Definition of Weights}
\label{sec:weight_method}

A key feature of the DREAMS suites is that they include MW-mass galaxies simulated under different combinations of cosmological and astrophysical parameters. However, not all parameter combinations will produce galaxies that are consistent with observations.
To address this, we assign weights throughout the parameter space based on how well the population of MW-mass galaxies simulated with those parameters reproduces the observed SMHM relation.

This method calculates weights for combinations of parameters, $\boldsymbol{\theta}$, rather than for individual galaxies.  
Even with a reasonable choice of parameters, an individual galaxy may not fall near some observed scaling relation because of e.g., a unique formation history or recent merger.  Sampling over a large number of galaxies---all generated with the same parameters---should give a better indication of how well those parameters reproduce the scaling relation.  

The CDM DREAMS suite is generated by varying $\Omega_{\rm m}$, $\sigma_8$, $\widebar{e}_w$, $\kappa_w$, $\epsilon_{f,{\rm high}}$.  When determining the weights, we marginalize over the two cosmological parameters because they have a minimal effect on the central galaxy properties compared to the astrophysical ones.  For the remaining three parameters, we consider an evenly-spaced $30\times30\times30$ grid that results in $2.7\times 10^4$ unique combinations. We choose the number of bins empirically to capture the detailed structure of the weights within the multidimensional parameter space while maintaining computational feasibility. However, the results do not change significantly as long as at least $\sim$10 bins are used per parameter.

The parameter combination for bin $j$ is indicated by $\theta_j$.  For each $\theta_j$, we generate $N_{j} =10^3$ emulated samples following the procedure outlined in Section~\ref{sec:emulator} and described in Appendix~\ref{app:nehod}.
This provides a stellar mass and halo mass for each galaxy in the bin, indicated as $X_{j,n}^{\rm em}$ and $Y_{j,n}^{\rm em}$ for the $n^{\rm th}$ emulated galaxy, respectively.
The number of generated samples, $10^3$, is chosen to ensure that the SMHM relation in each bin converges; see Figure~\ref{fig:sim_count}.

To assign a weight to each bin, we compare the emulator results with an observed SMHM relation.  To facilitate the comparison, the observed relation is fit with a piecewise linear function $g(X)$, where $X$ is the halo mass, and the slope and intercept can depend on $X$.  Then, for a given bin, we compare the SMHM relation for the emulated sample with the observed scaling relation by calculating the mean residual
\begin{equation}
    R_{\theta_j}\ =\ \frac{1}{N_j}\sum_{n=1}^{N_j} \left(Y_{j,n}^{\mathrm{em}}-g\left(X_{j,n}^{\rm em}\right) \right) \,.
\end{equation}
If the parameter combination $\theta_j$ produces galaxies that, on average, yield the observed scaling relation, $R_{\theta_j}$ will be close to 0. Systematic deviations from the relation will result in non-zero residuals.

This residual is converted into an unnormalized weight, $\widetilde{w}_j$, for the $j^{\rm th}$ bin using a Gaussian kernel with a free `temperature' parameter, $\tau$, as follows:
\begin{equation}
    \widetilde{w}_j = \mathrm{exp} \left( - \frac{R_{\theta_j}^2}{2 \tau^2} \right) \,.
\end{equation}
The final normalized weight for this bin is then
\begin{equation}
    w_j=\frac{\widetilde{w}_j}{\sum_j \widetilde{w}_j } \, ,
\end{equation}
where the sum in the denominator is taken over all the bins in the grid. 
The temperature parameter, $\tau$, controls how sharply to penalize deviations from the observed relation; effectively, it represents the uncertainty in the target observational mean relation.
As $\tau$ decreases toward zero, the weighted SMHM from the simulations will match the single parameter combination that minimizes the residual.
In this case, the contribution from parameter uncertainty to the scatter in the simulated SMHM disappears.
We choose $\tau$ to be 0.2 dex based on the uncertainties on the mean observed relation reported in~\cite{2019Behroozi}.
While~\cite{2019Behroozi} quotes a potential systematic uncertainty of $\sim$3~dex, the scatter between various observed constraints at the MW-mass scale is tighter, $\sim$1~dex (see their Figure~34).
We therefore select 0.2~dex as a representative value that accounts for the spread in the observed mean SMHM relations while acknowledging potential systematic offsets.
We find that our main results are robust to small variations in $\tau$, e.g., $\tau=0.2\pm0.1$~dex.

\begin{figure*}
    \centering
    \includegraphics[width=0.8\textwidth]{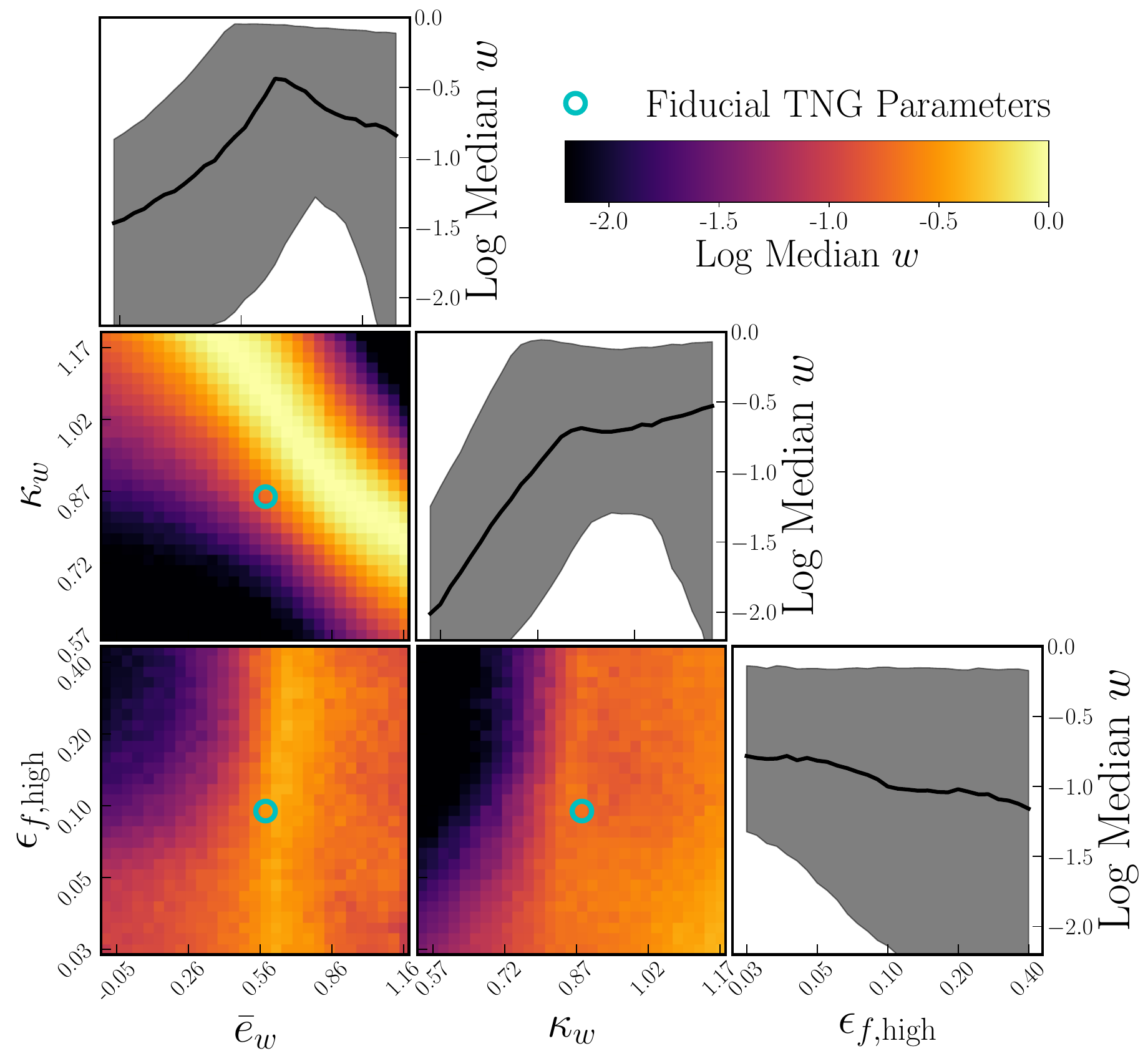}
    \caption{Corner plot showing the weights associated with the astrophysical parameters in the emulated DREAMS CDM dataset.  Each 2D panel shows the median weight, as defined in Section~\ref{sec:weight_method}, for a pair of parameters, marginalized over the other one. The highest weights are indicated by yellow/orange, while the lowest weights are indicated by blue/black.  
    The fiducial TNG values are shown with a cyan `o'. The top panel of each column shows the marginalized weight distribution for a given parameter, with the black line corresponding to the median and the gray band to the 16--84\% range. This procedure down-weights regions of parameter space that do not produce the data-driven SMHM relation from~\cite{SAGAV}.  Notably, there are significant degeneracies between some of the parameters, which could be further constrained by including other observations into the weighting procedure.}
    \label{fig:combined_weights}
\end{figure*}

This weighting procedure is deliberately lightweight, requiring only the ability to (i)~fit a linear regression on observed data and (ii)~simulate batches of $(X,Y)$ pairs from the model at different $\boldsymbol{\theta}$ values.
Alternative methods implemented in various SBI analyses, such as approximate Bayesian computation~\citep{lintusaari2017fundamentals} or neural density estimation~\citep{2021Papamakarios}, require carefully tuned discrepancy measures, ad hoc tolerances, or computationally expensive estimators that may obscure the structure of the parameter-data relationship.
This method acts as a simple and theoretically grounded alternative that provides a representation of how well $\boldsymbol{\theta}$ reproduces the empirically observed relationship between covariates $X$ and outcomes $Y$ through an auxiliary linear model fit on real data. 

Figure~\ref{fig:combined_weights} shows the median weights for each $\boldsymbol{\theta}$, projected in two-parameter planes. When projected into two dimensions, each plane marginalizes over the other three parameters, with the median weight of each bin shown.
Areas of high weight are indicated in yellow/orange, while areas of low weight are indicated in blue/black.  
Within each 2D projection, the cyan `o' indicates the fiducial TNG values.
The top panel of each column shows the marginalized distribution for each parameter, with the black line indicating the median and the gray band indicating the 16--84\% spread.
The cosmological parameters are not shown as they do not significantly affect the SMHM relation from the DREAMS simulations and are not included in the weight calculation.

\begin{figure*}    
    \centering
    \includegraphics[width=\textwidth]{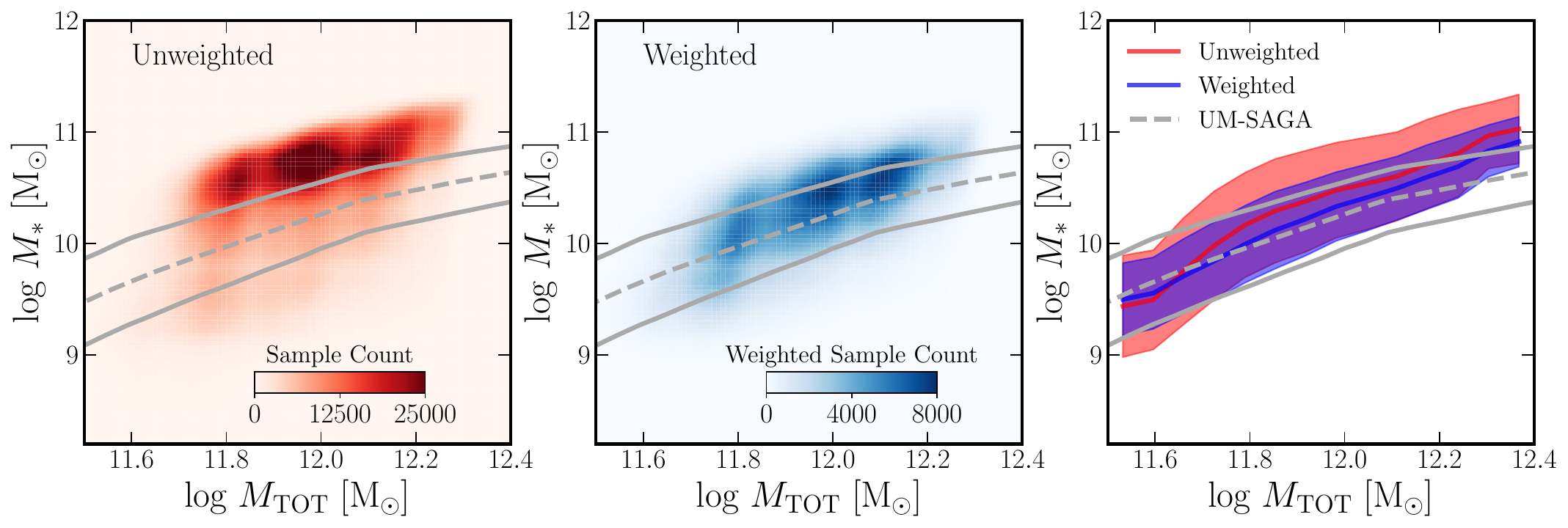}
    \caption{The SMHM for the emulated DREAMS CDM galaxy population, used to constrain the astrophysical parameters \kw, \ew, and $\epsilon_{f,{\rm high}}$.
    The left panel shows the full, unweighted distribution of emulated galaxies as a 2D histogram, with color indicating the density of points.
    The center panel shows the weighted distribution, after applying the scheme from Section~\ref{sec:weights} to down-weight unphysical regions of parameter space.
    The right panel directly compares the unweighted~(red) and weighted~(blue) populations, showing their mean (solid lines) and 1$\sigma$ intrinsic halo-to-halo scatter~(shaded bands).
    In all three panels, the UM-SAGA SMHM relation from~\cite{SAGAV} is shown for comparison as a dashed grey line~(mean) and solid gray bands~(1$\sigma$ scatter).
    Applying the weights brings the mean of the simulated population into closer agreement with observations.
    }
    \label{fig:unweighted}
\end{figure*}

The two SN parameters, \ew\,and $\kappa_w$, exhibit the most variation when varied individually, with \ew\,peaking near the TNG fiducial value and \kw\,plateauing above the TNG fiducial value.
The AGN parameter, $\epsilon_{f,{\rm high}}$, does not show as much variation but is still constrained by the SMHM relation at its lowest values.  This is likely due to the fact that, in the TNG model, MW-mass halos are not as dominated by AGN feedback as more massive galaxies.
The two cosmological parameters, \om\,and $\sigma_8$, show no variation when varied individually.

It is not surprising that each parameter individually peaks near the fiducial TNG value, as validation of the model often varied one parameter at a time~\citep{2018Pillepicha}.
As shown in Figure~\ref{fig:combined_weights}, though, the weights can be correlated between the different astrophysical parameters. 
This is especially evident for the two SN parameters, where the distribution peaks at higher values of \ew\,and lower values of \kw, or vice versa.  
In this case, it is strongly disfavored to simultaneously choose \ew\,and \kw\,at the min/max of their range.
The highly favored band of \ew\,and \kw\,is offset from the fiducial TNG value due to the higher resolution of the DREAMS simulations, which leads to higher stellar masses at a fixed halo mass~\citep{2018Pillepicha}.
This reflects that, within the TNG model framework, stronger SN feedback is required at the DREAMS mass resolution to regulate stellar masses to observed values.
These results highlight that sub-grid parameters can have correlated and/or non-linear impacts on model outputs. They underscore the benefit of exploring high-dimensional parameter spaces directly, rather than collapsing the problem to a series of one-parameter variations that can miss these complex interdependencies.

A significant implication of our weighting analysis is the non-uniqueness of the fiducial IllustrisTNG parameters.
The original TNG model, shown as `o' markers in Figure~\ref{fig:combined_weights}, lies within the extended region of high-weight parameter space for all five parameters.
This demonstrates that for the SMHM relations, there is no `correct' set of parameters required to produce realistic galaxies.
The large region of reasonable parameter space highlights the need for a DREAMS-like approach, which incorporates uncertainties on the underlying galaxy-formation model that are otherwise ignored in a traditional analysis.

The weighting procedure presented here can be further enhanced by adding additional observational inputs. For example, the SMHM relation used here is measured at $z=0$, but this analysis could be extended to higher redshifts, similar to what has been done for other simulations~\citep{2014Torrey, 2015Schaye, 2018Pillepicha}.
Additional present-day scaling relations could also be added, such as the Tully-Fisher relation~\citep{1977Tully,2019Lelli}, the mass-size relationship~\citep{2003Shen,2021Nedkova}, or the stellar mass--age relationship~\citep{2005Gallazzi,2017Goddard}.
Additionally, adding BH-specific scaling relations, like the $M\text{--}\sigma$ relation~\citep{2020Greene}, could better constrain \agn.
The analysis extends straightforwardly to the inclusion of these inputs. We save a more detailed exploration of this to future work.  

\subsection{Comparison of SMHM Relations}
\label{sec:unweighted}

This subsection compares the emulated DREAMS data to an observational SMHM relation, which describes the efficiency with which halos convert baryons to stars. In particular, we use the `UM-SAGA' model from Figure~5 in~\citet{SAGAV}, which updates the data-driven \textsc{UniverseMachine}~\citep[UM;][]{2019Behroozi} framework for modeling the galaxy-halo connection with the third data release from the Satellites Around Galactic Analogs Survey~\citep[SAGA;][]{SAGAI,SAGAIII,SAGAV}.
This updated SMHM relation agrees well with the original UM result~\citep{2019Behroozi,2021Wang} and other recent estimations~\citep{2020Nadler, 2018Pillepichb}.

For a fair comparison to the UM-SAGA model, we must ensure that our definitions of stellar and total halo mass are comparable to those used by~\cite{SAGAV}. For each DREAMS galaxy, we take $\mstar$ to be the mass of the stellar material bound to the central galaxy within $2r_\mathrm{*,half}$ at $z=0$, where $r_\mathrm{*,half}$ is the half-mass radius of all gravitationally bound star particles.
This is roughly equivalent to the stellar mass that is used in~\cite{SAGAV} as $\mstar$ includes all stars formed both \emph{in situ} and \emph{ex situ} and does not include satellites orbiting the host at $z=0$.
For the total halo mass, $M_{\rm TOT}$, we take the sum of all mass within $R_{99.2}$, the radius where the density of the galaxy is $99.2$ times the critical density to be consistent with~\cite{SAGAV}.
We do not use the peak halo mass, as is done in~\cite{SAGAV}; however, nearly all DREAMS central galaxies obtain their peak mass at $z=0$. Approximately 10\% of simulations have a peak mass that is up to 3\% higher than their $z=0$ total mass, and $\sim$1\% have a peak mass that is $>10$\% larger than their $z=0$ total mass.
Therefore, we do not expect this difference to affect the results presented here.

These mass definitions provide an even-handed comparison between the  DREAMS galaxies and the UM-SAGA model. The left and middle panels of Figure~\ref{fig:unweighted} show the 2D histogram for the unweighted and weighted emulated DREAMS data, respectively.  For comparison, the mean of the UM-SAGA model is shown by the dashed gray line, with the $1\sigma$ band indicated by the solid gray lines.  The right-most panel of Figure~\ref{fig:unweighted} directly compares the  results by plotting the mean and $1\sigma$ band for the unweighted and weighted distributions.  In general, the unweighted distribution is shifted above the UM-SAGA model, but weighting the emulated data brings it closer to the observed relation. 

However, it is worth noting that the weighted result does not agree identically with the UM-SAGA model, which may be due to several reasons, beyond the choice of $\tau$.  First, there may be systematic differences between how properties are measured in simulations versus how the observations are performed. 
Second, there may be other uncertain parameters in the TNG model beyond those varied here and/or missing physics beyond the TNG sub-grid model. 
This motivates future simulation suites that vary over a larger number of sub-grid parameters, such as was done by~\cite{SB28} for uniform-box simulations, as well as exploration of alternative sub-grid models to TNG.  
Additionally, we note that the calculated parameter weights will depend on resolution, as the stellar mass of the host galaxies increases with resolution~\citep{2018Pillepicha}.



The exponential form of the weighting function ensures that parts of parameter space producing unrealistic galaxies are penalized and provides a physically motivated guide for understanding the galaxy-formation model's dependencies.  In the following sections, we apply these weights to all results so that the uncertainty on measured MW properties is representative of the current uncertainty from the astrophysical parameters.
We have also repeated the analyses in this work without using the weights and found that the overall conclusions remain unchanged.

\section{Milky Way Properties}
\label{sec:mw_props}

\begin{figure*}
    \centering
    \includegraphics[width=\textwidth]{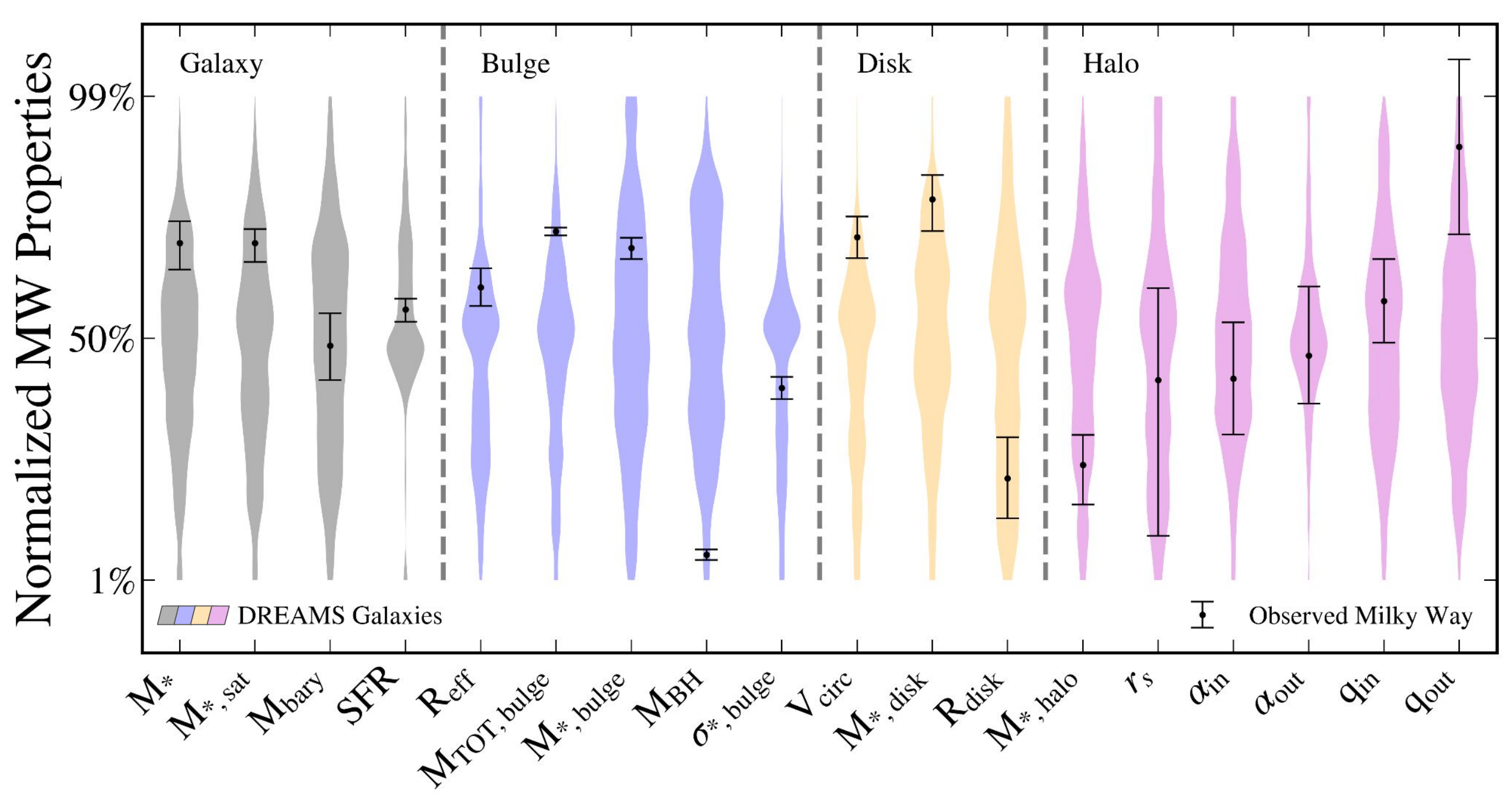}
    \caption{A summary of the key physical properties of the MW-mass galaxies in the DREAMS CDM suite.  The value for each property is normalized and clipped to the $1^{\rm st}$ and $99^{\rm th}$ percentile of the simulated values, see Table~\ref{tab:props} for the unnormalized values.
    The weighted distributions for the 803 DREAMS disk galaxies are shown as colored violins, grouped by galactic~(gray), bulge~(blue), disk~(yellow), and halo~(pink) properties.  Note that these results correspond to the actual simulated galaxies, not the emulated set shown earlier. 
    Detailed descriptions for each property can be found in Section~\ref{sec:props}.
    The black points with 1$\sigma$ error bars show the corresponding observed values and their uncertainties for the MW taken from~\cite{2016Bland}.  These results put our Galaxy in context, providing a direct comparison with the broader simulated population in the DREAMS suite.  }
    \label{fig:violin}
\end{figure*}

This section explores the various properties of the central MW-mass galaxies in the DREAMS CDM suite.
By combining the statistical power of these 1,024 unique systems with the physically motivated uncertainties derived from the parameter weighting, this suite provides a powerful tool for contextualizing our Galaxy.
Note, however, that the isolation criteria imposed on the DREAMS halos restrict them to lower-density environments (see Section~\ref{sec:sims}) and do not allow for a M31 analog.

Figure~\ref{fig:violin} shows distributions of the galactic~(grey), stellar bulge~(blue), stellar disk~(yellow), and stellar halo~(pink) properties of the DREAMS simulations.  These are compared with data from observations, which are indicated by the black points and taken from~\citet{2016Bland}.
The violins in Figure~\ref{fig:violin} correspond directly to the simulations and not the emulated data.  
Each galaxy in the DREAMS CDM suite is assigned a weight based on its astrophysical parameters, and the violins in Figure~\ref{fig:violin} are the weighted distributions of simulated properties.

The distributions presented in Figure~\ref{fig:violin}, and later presented in Section~\ref{sec:gse}, reflect the total uncertainty in our predictions, which arises from three distinct components.
First, the suite covers a broad range of virial masses, $(0.5\text{--}2.0)~\times~10^{12}~\mdot$, to account for the observed uncertainty in the MW's true host halo mass.
This introduces a systematic spread in the predicted value of galactic properties that scale with halo mass.
Second, we explicitly vary five cosmological and astrophysical parameters (\om, \s8, \ew, \kw, \agn).
While we have placed constraints on these parameter combinations in Section~\ref{sec:weights}, uncertainty remains in the correct choice of values.
Finally, intrinsic halo-to-halo variance plays a significant role as we sample 1,024 unique initial density fields across the DREAMS simulation suite, resulting in 1,024 unique merger histories and environments.
We focus this analysis on the relative contribution of the five simulation parameters, as DREAMS is uniquely suited to probe these parameter variations, but future work can further investigate the relative impact of halo mass uncertainty and intrinsic halo-to-halo variance.
Below, we briefly overview each of the properties used for this comparison.

\subsection{Measured Properties}
\label{sec:props}

For each galaxy in the DREAMS CDM suite, we identify the simulation particles that are gravitationally bound to it and located within $R_{200}$, excluding those bound to resolved satellites. 
To aid in the decomposition, we define \emph{ex-situ} stars as those star particles that are further than 50~ckpc from the center of the galaxy at formation time, and \emph{in-situ} stars as those that remain. This spatial cut is motivated by prior work demonstrating that stars formed at such large distances typically make up the bulk of the stellar halo~\citep[e.g.,][]{2015Pillepich}. To calculate this distance, we track the main progenitor branch back in time to $z=15$ and interpolate the position of the potential minimum at the time of each star's birth using a cubic spline.

We decompose the stellar component of each galaxy into its primary structural components---disk, bulge, and halo---using the following criteria:
\begin{itemize}
\item \textbf{Disk stars} are selected stochastically from co-rotating \emph{in-situ} stars. 
First, we identify \emph{in-situ} stars throughout the galaxy that are counter-rotating compared to all stars between 1--5~kpc.
We assume that there is a symmetric co-rotating \emph{in-situ} component.
Disk stars are then chosen stochastically, based on their circularity parameter $\epsilon$, where
\begin{equation}
    \epsilon = \frac{L_z}{L_z^{\rm max}(E)} ,
\end{equation}
$L_z$ is the $z$-component of the angular momentum, and $L_z^{\rm max}(E)$ is the maximum angular momentum that is allowed for the specific orbital energy. 
From the measured $\epsilon$ for each \emph{in-situ} star particle, the probability of the star being part of the disk is estimated by dividing the total $\epsilon$ distribution by the symmetric spheroidal $\epsilon$ distribution, consistent with the method outlined in~\cite{2023Orkney}.
\item \textbf{Bulge stars} are then selected as the remaining stars, both \emph{in-situ} and \emph{ex-situ}, that pass our binding energy criteria.
We require bulge stars to be more tightly bound than a star particle traveling in a circular orbit at 3~kpc. 
This reference radius was determined empirically by varying the cutoff; 3 kpc marks the transition where the enclosed mass profile flattens, distinguishing the dense bulge from the diffuse stellar halo.
\item \textbf{Halo stars} are identified as the remaining \emph{in-situ} and \emph{ex-situ} stars that are not selected to be part of the disk or bulge (i.e., loosely bound \emph{ex-situ} stars and \emph{in-situ} stars that are both loosely bound and not rotating within the disk).
\end{itemize}


We find that 22\% of the DREAMS galaxies do not host a resolved disk and thus cannot be easily separated into bulge, disk, and halo components.
Specifically, these galaxies do not contain at least $5\times10^3$ star particles that would be selected as part of the disk outside of 3~kpc.
Most of the unresolved galaxies reside in regions of parameter space with high SN feedback (high \ew\,and/or high \kw).
Additionally, we remove 22 galaxies that have severely warped disks due to a recent merger.
Since the decomposition of these galaxies is ambiguous and does not match the morphology of the MW, they are not included in the remaining analyses of this paper, resulting in 803 disk galaxies analyzed in this section and Section~\ref{sec:gse}.

The criteria for distinguishing halo, bulge, and disk stars are designed to match MW observations as closely as possible.
In observations, halo stars are typically identified by having significant velocity dispersions, non-circular orbits, and low metallicities~\citep{2007Carollo,2010Carollo}.
While the best match to this selection function would be to track the stellar orbits and metallicities, the stellar halo metallicities vary throughout the halo, and a consistent metallicity criterion that fits all 1,024 galaxies with different feedback models may not be possible~\citep{2005Bullock,2018Belokurov,2018Helmi,2019Conroy}.
Instead, we rely on kinematics and accretion to track which stars belong to the halo, which has been shown to accurately locate halo stars~\citep{2009Zolotov,2011Font,2019Monachesi}.

The selection of disk stars through circularity is analogous to observational techniques that identify cold and rotationally supported material~\citep{2003Bensby}.
Bulge stars are observationally identified through a combination of morphology and kinematics~\citep{2010Minniti, 2012Kunder, 2013Wegg}, which are reasonably approximated by our selection of centrally bound stars that are not part of the disk or halo.
The MW's bulge is expected to be dominated by a pseudo-bulge that is formed through secular processes. At most $\sim$$10\text{--}25\%$ of the mass is expected to be in the pressure-supported classical bulge~\citep{2010Shen,2019Clarke} formed through mergers.
Interestingly, 77\% of disk galaxies in the DREAMS simulations have at most 25\% of their bulge mass in \emph{ex-situ} stars, making most consistent with the MW in this regard.

\begin{table*}
    \centering
    \begin{tabular}{l|lll|ll|ll}
        \Xhline{3\arrayrulewidth}
        \textbf{Property} & \textbf{1\%} & \textbf{50\%} & \textbf{99\%} & \textbf{Observed Value} & \textbf{Observed Uncertainty} & \textbf{Units} & \textbf{Log} \\
        \hline
        $\mathrm{M_*}$        &  8.97  & 10.54  & 11.31 & 10.70  & [10.60, 10.78] & $\mdot$  &  Yes  \\
        $\mathrm{M_{*,sat}}$  &  5.28  &  8.77  & 10.80 &  9.43  & [9.28, 9.54]  & $\mdot$  &  Yes  \\
        $\mathrm{M_{bary}}$   & 10.39  & 10.98  & 11.48 & 10.93  & [10.87, 10.99] & $\mdot$  &  Yes  \\
        SFR                   & $-4.00$*  & $-0.37$  &  1.71 &  0.22  & [0.16, 0.27]  & $\mdot~\mathrm{yr}^{-1}$  &  Yes  \\
        \hline
        $\mathrm{R_{eff}}$        &  0.24   &   0.56 &   4.99 &  0.70 & [0.63, 0.77]  & kpc          &  No  \\
        $\mathrm{M_{TOT,bulge}}$  &  9.12   &  10.05 &  10.99 & 10.27 & [10.25, 10.28] & $\mdot$      &  Yes  \\
        $\mathrm{M_{*,bulge}}$    &  8.43   &   9.80 &  10.96 & 10.19 & [10.15, 10.23] & $\mdot$      &  Yes  \\
        $\mathrm{M_{BH}}$         &  6.28   &   7.60 &   8.70 &  6.62 & [6.60, 6.64]  & $\mdot$      &  Yes  \\
        $\sigma_\mathrm{*,bulge}$ & 54.25   & 186.29 & 536.56 & 113.0 & [110.0, 116.0] & km s$^{-1}$  &  No  \\
        \hline 
        V$_\mathrm{circ}$     & 96.43   & 197.52 & 385.61 & 238.00 & [223.00, 253.0] & km s$^{-1}$  &  No    \\
        $\mathrm{M_{*,disk}}$ &  7.95   &  10.22 &  11.16 &  10.61 & [10.48, 10.71] & $\mdot$      &  Yes    \\
        $\mathrm{R_{disk}}$   &  1.09   &   4.32 &  14.58 &   2.51 & [2.06, 2.97]  & kpc          &  No    \\
        \hline
        $\mathrm{M_{*,halo}}$  & 7.95    &  9.28 & 10.38 &  8.74 &  [8.60, 8.85] & $\mdot$  &  Yes \\
        $r_s$                  & 9.08    & 29.25 & 63.10 & 25.00 & [15.0, 35.0]  & kpc  &  No  \\
        $\alpha_\mathrm{in}$   & $-4.84$   & $-2.32$ &  0.89 & $-2.5$  & [$-2.8$, $-2.2$]  & N/A  &  No  \\
        $\alpha_\mathrm{out}$  & $-12.19$  & $-4.19$ & $-0.74$ & $-4.35$ & [$-5.0$, $-3.7$]  & N/A  &  No  \\
        q$_\mathrm{in}$        & 0.18    &  0.62 &  0.95 &  0.65 & [0.60, 0.70] & N/A  &  No  \\
        q$_\mathrm{out}$       & 0.18    &  0.59 &  0.89 &  0.80 & [0.70, 0.90] & N/A  &  No  \\
        \Xhline{3\arrayrulewidth}   
    \end{tabular}
    \hspace*{1.5cm}
    \caption{This table provides the unnormalized values corresponding to the violin plots in Figure~\ref{fig:violin}. 
    Columns show the 1$^{\rm st}$, 50$^{\rm th}$ (median), and 99$^{\rm th}$ percentile of the 803 DREAMS disk galaxies plotted in Figure~\ref{fig:violin}.
    The observed MW values and their $1\sigma$ uncertainties from~\cite{2016Bland} are included for comparison and  correspond to the black points in Figure~\ref{fig:violin}.
    The last column indicates whether the quantities are provided as log of the physical units shown in the `Units' column.
    *A value of $-4$ for the SFR is below the resolution of the DREAMS simulations, but is used to represent galaxies that are quenched on the log scale.
    }
    \label{tab:props}
\end{table*}

For each of the structural components specified above, we focus on a few key structural parameters.  
$\mstar$\footnote{Note that this is a slightly different definition from the one used in Section~\ref{sec:weights}.} is the total mass of all star particles in the galaxy, while $M_{*, {\rm sat}}$ is the mass of the stellar material bound to its satellites. 
The total baryon mass, $M_{\rm bary}$, is the total mass of all galactic stars and gas, including both hot and cold components, and excluding satellites to match the observational measurement~\citep{2016Bland}. 
The host's SFR is the summed instantaneous star-formation rate of the gas in the galaxy at $z=0$.

The scale radius for the bulge surface-density profile, $R_{\rm eff}$, is not uniform for the MW due to the box/peanut shape of its bulge~\citep{2013Wegg}.
Since this is also the case for many of the DREAMS galaxies due to both asymmetries and bars, we measure cylindrically averaged stellar-density profiles along 500 unique lines of sight and take the largest $R_{\rm eff}$.
The lines of sight are spread uniformly over a Fibonacci sphere~\citep{fibonacci} to ensure all angles are sampled.
For each line of sight, the surface-density profile is measured over a cylinder with a length of 10~kpc and a minimum diameter of 500~pc. 
We require a minimum difference of the three largest measured $R_{\rm eff}$ to be within 0.5~kpc of each other to ensure that the measurement is robust to numerical artifacts.
This method ensures that $R_{\rm eff}$ is measured along bar-like features to match  observations~\citep{2016Bland}.

$M_{\rm TOT,bulge}$ is the total mass of all material in the bulge (including all stars, gas, DM, and BHs) within $3R_{\rm eff}$. It acts similarly to a dynamical mass. The size of the region, $3R_{\rm eff}$, roughly matches the Vista Variables in the Via Lactea survey~\citep[VVV;][]{2010Minniti} region of our Galaxy where the corresponding observational quantity is measured~\citep{2016Bland}.
$M_{*, {\rm bulge}}$ is the summed mass of star particles in the bulge.
$M_{\rm BH}$ is the mass of the single SMBH particle found at the minimum of the host potential.
As a reminder, seed BHs of $1.2\times10^{6}~\mdot$ are populated in halos that exceed a mass of $7.2\times10^{10}~\mdot$ in the TNG model~\citep{2017Weinberger,2018Weinberger}.
Finally, the total velocity dispersion of the stars in the bulge, $\sigma_\mathrm{*, bulge}$, is the standard deviation of the bulge star speeds.

The disk properties include the circular velocity, $V_{\rm circ}$, derived from the spherically averaged gradient of the potential at 8~kpc from the center of the host.  This property will likely be correlated with $M_{*, {\rm bulge}}$, $M_{\rm TOT,bulge}$, and the stellar mass of the disk, $M_{*,{\rm disk}}$. 
We determine the disk radial scale length, $R_{\rm disk}$, by fitting an exponential profile, $\Sigma(R) \propto \text{exp}(-R/R_{\rm disk})$, to the stellar surface-density profile. The fit is performed in the range $3<R<30$~kpc, after aligning the coordinate system with the disk's angular momentum.\footnote{Our results are not sensitive to the choice of maximum radius 30~kpc. The method of decomposing the bulge, disk, and halo stars results in good exponential fits to the surface-density profile out to large radii, such that using a smaller maximum radius does not change the results.} 
Given that the DREAMS MWs contain $\sim$$10^6$ DM particles, there should not be a significant ($<10\%$) artificial increase to the size of the disks from spurious collisional heating~\citep{2021Ludlow}. However, we do not resolve the thin- and thick-disk components due to a spatial softening of 0.441~kpc.  As a result, we take the observed value of $R_{\rm disk}$ to be a mass-weighted average of the thin- and thick-disk radial scale lengths from~\cite{2016Bland}.


The mass of the stellar halo, $M_{*, {\rm halo}}$, is the sum of all star particles selected as being part of the halo and that have a galactocentric radius of 10--45~kpc to match the observational selection~\citep{2016Bland}.
The break radius, $r_{\rm s}$, is found by fitting a double power law~(DPL) to the cylindrically averaged surface-density profile.
The slope of the inner power law, $\alpha_\mathrm{in}$, and outer power law, $\alpha_\mathrm{out}$, are also measured from the same DPL fit.
We find that 17\% of stellar halos have an inner and outer slope within 0.5 of each other and are better fit by a single power law rather than a DPL.
For these halos, we do not measure a break radius as its exact location is ambiguous.
Finally, the shape of the halo is measured as the inner, $q_{\rm in}$, and outer, $q_{\rm out}$, ellipsoidal flattening of the halo stars in the corresponding regions defined by the DPL fit, where $q \sim 1$ is spherical and $q \sim 0$ is elongated.
The flattening is defined as the ratio of the short-to-long axes of a three-dimensional ellipsoidal fit to the mass-weighted stellar distribution.

\subsection{The Milky Way in Context}
\label{sec:context}

Figure~\ref{fig:violin} shows the distribution of DREAMS galaxies for each property described in Section~\ref{sec:props} as violins centered on the median and extending to the 1\% and 99\% percentiles of the data.  
Observational values and their uncertainties are taken from~\cite{2016Bland} and are displayed by the black points.  
The value for each property is normalized such that the median is mapped to 0 (50\%), the $1^{\rm st}$ percentile to $-1$~(1\%), and the $99^{\rm th}$ percentile to 1~(99\%).
For reference, the unnormalized values are provided in Table~\ref{tab:props} and Figure~\ref{fig:1D_props}.

The MW stands out against the larger population probed by the DREAMS galaxies for some of the structural properties in Figure~\ref{fig:violin}.  For example, $M_{*, {\rm sat}}$ is higher for the MW than the median DREAMS galaxy; this is driven largely by the LMC, which makes up 75\% of the stellar mass of all satellites around the MW~\citep{2020Nidever,2021Shipp,2024Watkins,2024Pace}.
Other properties, such as the MW's baryon mass, $M_{\rm bary}$, are typical compared to the expectation of the DREAMS suite. 
The observed SFR of the MW is near the DREAMS median.  
Interestingly, a $\sim$10\%  fraction of the DREAMS galaxies are quenched at $z=0$, as indicated by the increase in the distribution near the 1\% limit.
However, these simulations are heavily disfavored by our weighting scheme, due to their deviation from the observed SMHM relation and thus appear as a small population.

The MW's galactic bulge is close to the average size of the DREAMS galaxies, measured by $R_\mathrm{eff}$.
Most of the DREAMS galaxies with bulge sizes near the MW host a bar-like feature.
From inspecting the surface-density profiles that $R_\mathrm{eff}$ is measured from, DREAMS galaxies with very large $R_\mathrm{eff}$ have a spherical bulge but are flattened in the center where the measurement is taken, and the long tail of low $R_\mathrm{eff}$ galaxies are galaxies with exponential bulges without bars.
The stellar and total mass of the bulge, $M_{*, {\rm bulge}}$ and $M_{{\rm TOT}, {\rm bulge}}$, are greater than $\sim$$70$\% of the DREAMS simulations but still consistent with the bulk of the distribution.
The largest outlier is the SMBH at the center of the galaxy, at $4.3\times10^6~\mdot$~\citep{2022GRAVITY}, which is less massive than 94\% of the DREAMS galaxies.
As noted earlier, the large seed mass already comprises $\sim$$25\%$ of the mass of Sagittarius~A*~\citep{2016Bland}, creating a high mass floor for the simulated BHs.
Similarly, the stellar velocity dispersion of the bulge, $\sigma_\mathrm{*,bulge}$, which is correlated with $M_\mathrm{BH}$, is also below the median DREAMS velocity dispersion.

The bulge identification used in this work does not distinguish between a classical dispersion-supported bulge and a pseudo rotationally-supported bulge. 
One might worry that this may lead to pseudobulge stars being misidentified as disk stars. 
To test whether this affects our results, we also implement a full kinematic decomposition based on the method of~\cite{2022Zana}, which separately identifies the two bulge components, for four properties, $M_\mathrm{*,bulge}$, $R_{\rm eff}$, $M_{\rm *,disk}$, and $R_\mathrm{disk}$.  
We find no significant differences in the global distributions, with the means remaining within a few percent of each other. 
The standard deviation about the mean varies by $\pm 20$\% in all cases except for $R_{\rm eff}$, where it decreases by a factor of 2.
As a result, using the method outlined in \cite{2022Zana} does not change the general conclusions drawn when comparing the DREAMS data to observations.  

The MW's disk is exceptional in that it is both more massive than the average DREAMS disk and more radially compact.
The unusual compactness of our Galaxy's disk has been noted in prior works~\citep[e.g.,][]{Hammer07,Bovy13,Licquia16a,Boardman20,Tsukui25}, although see also~\citet{Lian24}.
Compared against the DREAMS galaxies, the MW's disk is more massive than $\sim$76\% and more radially compact than $\sim$78\% of the disk galaxies in the DREAMS suite.
The MW's massive and compact disk may influence the Galaxy's structure from the inner DM halo to orbiting satellites, as suggested by e.g.,~\citet{2004Gnedin,2007Hammer,2008Debattista,2016Licquia,2017Garrison-Kimmel,2020Boardman,2025Tsukui}.

While the uniqueness of the MW's stellar halo is less apparent due to large observational uncertainties, it has two characteristics that stand out.
First, its $M_{*, {\rm halo}}$ is lower than 89\% of the DREAMS galaxies, and second, its outer halo is more spherical than 94\% of the DREAMS galaxies.
The MW break radius and surface-density slopes are all consistent with the DREAMS galaxies, with the uncertainties covering a significant fraction of the spread within the DREAMS results.

Overall, the DREAMS galaxies provide an excellent range of properties to contextualize the MW and other similar galaxies.
A crucial success of this DREAMS suite is that the collective properties of the simulated galaxies, shown in Figure~\ref{fig:violin}, span the observed properties of the MW from its massive, compact disk to its undermassive SMBH.
Re-analyzing the simulations with baryonic physics fixed to fiducial TNG values, using an emulated sample (see Section~\ref{app:TNG}), results in different distributions for some properties, mostly limited to the bulge region.
Specifically, $M_{\rm bary}$, $M_\mathrm{TOT,bulge}$, $R_{\rm eff}$, $M_\mathrm{BH}$, $\sigma_\mathrm{*,bulge}$, and $V_{\rm circ}$ are affected.
While all these distributions shift, only $M_\mathrm{BH}$ has a tighter distribution with fixed physics, indicating that all other galactic properties are dominated by a combination of halo mass uncertainty and intrinsic halo-to-halo variance.

This demonstrates that the model is capable of producing MW analogs, even its more rare qualities like its small SMBH or massive disk, although there is no single galaxy that possesses all of the MW's unique characteristics simultaneously.
While other hydrodynamical simulation projects, like Auriga~\citep{2017Grand} and Apostle~\citep{APOSTLE}, have endeavored to reproduce the MW's more unique characteristics through suites of simulations which match the MW's mass or local environment, the simulations in the DREAMS suite span a larger range of the MW's characteristics, even if no single simulated galaxy is a perfect analog.
This provides a rich dataset for understanding how the MW's specific history shaped its present-day properties, including many beyond those that we present here.

\section{GSE Effects on Galactic Properties}
\label{sec:gse}

This section demonstrates how the DREAMS simulations can be used to investigate the origin of the MW's unique characteristics.
We explore the connection between the specific history of the MW, in particular its GSE merger, and present-day properties. Section~\ref{sec:gse_merger} defines the selection criteria for GSE-like events.  Then, Section~\ref{sec:gse_rarity} discusses how the free parameters of the simulation affect the rarity of a GSE-like event.  In cases where such an event occurs, Section~\ref{sec:gse_props} shows how properties of the MW host are likely to be affected.
Additional discussion and analysis are provided in Appendix~\ref{app:nehod} to validate the emulator results used in this section.

\subsection{The GSE Merger and its Analogs}
\label{sec:gse_merger}

With \textit{Gaia}'s precise measurements of local halo  stars, a radially anisotropic structure in the velocity distribution was definitively revealed and attributed to the GSE merger~\citep{2018Belokurov,2018Helmi}---for a review, see~\citet[][]{2020ARA&A..58..205H}.
It is believed that the GSE is the MW's last major merger, occurring approximately 8--10~Gyr ago with a stellar mass ratio ranging from 1\,:\,5 to 1\,:\,2 relative to the MW~\citep{2018Belokurov,2018Helmi,2019Gallart,2020Bonaca,2021Feuillet,2021Grunblatt,2021Naidu,2021Hasselquist,2024Horta}.
There is no evidence for another major merger since the GSE event;  the most noteworthy recent {\it minor} merger is Sagittarius\footnote{For the definitions we use in the GSE analysis, the LMC would not be considered to have started its merger with the MW yet since it is still star forming and has not undergone significant tidal stripping~\citep{2007Besla,2022Massana,2024Sheng}.}.

At the time of the GSE merger, the progenitor galaxy is suspected to have disrupted the MW's stellar disk, potentially leading to the formation of the present-day thick disk~\citep{2018Belokurov,2020Naidu,2022Belokurov,2025Woody}.
Additionally, the accreted stars from the GSE dwarf comprise a dominant component of the present-day inner stellar halo~\citep{2018Helmi,2018Deason,2018Helmi,2019Lancaster,2018Myeonga,2018Myeongb,2021Naidu,2022Han}.
One or two break radii are thought to be present in the MW's stellar halo~\citep{2011Deason,2012Kafle}, which may correspond to the first and second apocenters of the GSE merger~\citep{2013Deason,2019Deason}.
If so, further understanding the conditions of this merger can help us to understand both the present-day properties of our Galaxy and its past.

To find GSE analog events in the DREAMS galaxies, we utilize the \textsc{Sublink}~\citep{Sublink,Illustris} merger tree catalogs.
\textsc{Sublink} tracks halos through time by identifying them  in consecutive simulation snapshots.
For a given halo, its descendant in the subsequent snapshot is identified through a score that accounts for DM particles that are shared in both snapshots, prioritizing those that are deeper in the potential.
This method efficiently and robustly tracks the evolution of subhalos through time, allowing for the construction of detailed merger trees that record the full accretion history of each galaxy.

We use these merger trees to identify MW-mass galaxies with GSE analogs based on five criteria:
\begin{itemize}
    \item The merger must have a stellar mass ratio of at least 1\,:\,5 relative to the MW ($M_{\rm MW}/M_{\rm gse}<5$) at the time of accretion. Here, the time of accretion is defined as the last time that the progenitor has its peak mass, just before it loses mass due to interactions with the host galaxy.
    \item The merger must have occurred between $z\in [1, 2]$, which corresponds to $\sim$$8\text{--}10$~Gyr ago, although the exact timing depends on the specific cosmology of the simulation.  Defining $t_{\rm gse}$ as the scale factor for the accretion time, this requirement is $t_{\rm gse}\in [0.33, 0.5]$. 
    \item Since the GSE-like event, there can be no mergers larger than Sagittarius, whose mass ratio is no greater than approximately $\sim$$1$\,:\,50~\citep{2010Law,2021Vasiliev}.
    Defining $t_{\rm mm}$ as the scale factor associated with the time of the last merger with a stellar mass ratio of at most 1\,:\,50, this requires that  $t_{\rm mm}<0.5$.
    \item The $z=0$ stellar debris from the GSE-like merger must comprise at least 50\% of the stellar halo's \emph{ex-situ} stars located within $40$~kpc of the galactic center ($f>0.5$), which must also have a radial anisotropy, $\beta$, greater than 0.5.  The latter is defined as 
    \begin{equation}
    \beta = 1 - \frac{\sigma_\phi^2 + \sigma_\theta^2}{2 \sigma_r^2} \,, 
    \end{equation}
where each $\sigma$ is the stellar dispersion measured in a given spherical direction.
\end{itemize}
Together, these five criteria select galaxies that, like the MW, experienced a significant, early, and radial merger that was then followed by a quiet merger history.  Note that these selections include the case of multiple mergers contributing to the GSE-event \citep{2022Donlon}, see~\cite{2025Folsom} for a discussion.
In this scenario, the total stellar material from the progenitors is used to calculate $M_{\rm MW}/M_{\rm gse}$, $f$, and $\beta$. 
The timing parameter, $t_{\rm gse}$, is set to the accretion time of the progenitor with the largest $f$ contribution.
The same five criteria are then applied to candidate hosts to select them as a GSE analog.

Within the DREAMS CDM suite, it is uncommon to find hosts with quiescent merger histories like our own.
Only 18\% of the systems have not had a major merger (e.g., one with a stellar mass ratio of at least 1\,:\,50) in the last 8~Gyr ($t_{\rm mm}<0.5$).
Furthermore, only 7.5\% of the systems have not had a merger with a stellar mass ratio of at least 1\,:\,10 in the same time period.
Of all the disk galaxies in the DREAMS suite, only 1.5\%~(11 galaxies) have GSE analogs, defined using the five criteria above.  Adopting the higher-end estimates for the GSE's mass (e.g., a stellar mass ratio of 1\,:\,2), this fraction drops to only 0.8\%~(6 galaxies).                    

\subsection{The Rarity of GSE Analogs}
\label{sec:gse_rarity}

\begin{figure*}
    \centering
    \includegraphics[width=\columnwidth]{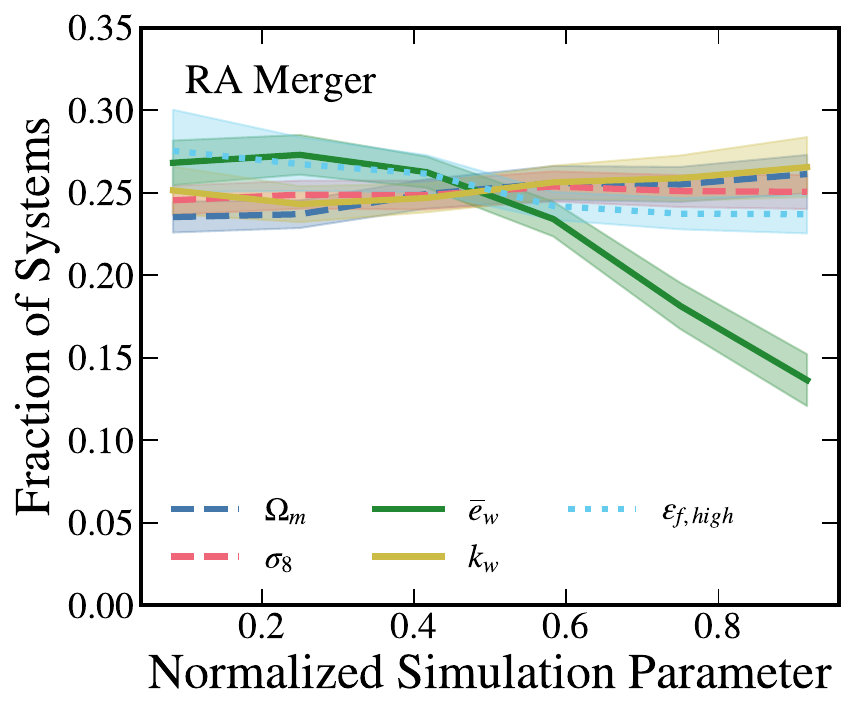}
    \includegraphics[width=\columnwidth]{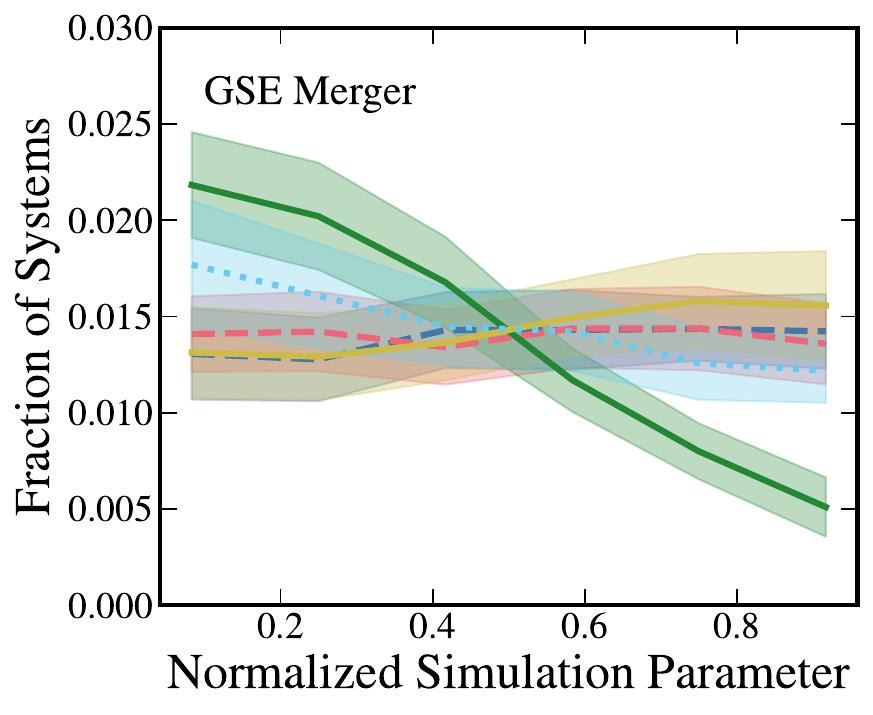}
    \caption{The fraction of emulated DREAMS hosts that undergo an RA event~(left) and a GSE-like event~(right), plotted as a function of the five simulation parameters ($\Omega_{\rm m}$, $\sigma_8$, $\widebar{e}_w$, $\kappa_w$, $\epsilon_{f,{\rm high}}$), normalized to unity.  
    The lines are the means of the emulated samples with the $1\sigma$ error bands corresponding to the aleatoric and epistemic uncertainties. Note the difference in scales between the y-axes of each figure panel.
    We adopt the definition of an RA event from~\cite{2025Folsom} where the debris from the merging galaxy makes up 50\% of the $z=0$ inner stellar halo and has a significant radial anisotropy, $\beta>0.5$.
    A GSE merger is an RA event, with additional criteria on the mass and accretion of the progenitor, as well as the host's accretion history (see Section~\ref{sec:gse_rarity}).  The occurrence of RA- and GSE-like events is most strongly correlated with the energy of SN winds, \ew. }
    \label{fig:gse_prob}
\end{figure*}

This subsection leverages generative machine learning models to investigate how the simulation parameters ($\Omega_{\rm m}$, $\sigma_8$, $\widebar{e}_w$, $\kappa_w$, $\epsilon_{f,{\rm high}}$) impact the frequency of GSE analog events.  To isolate the effect of each simulation parameter and generate a statistical sample of MW-mass galaxies with GSE analogs, we train a separate emulator with the same architecture as the one outlined in Section~\ref{sec:emulator}.
The general training process is the same as that described earlier, except that the target parameters are now different.  Specifically, they are set to the fraction of progenitor debris in the inner stellar halo~($f$), the stellar velocity anisotropy~($\beta$), the scale factor of the last major merger~($t_\mathrm{mm}$), the scale factor of the GSE-event~($t_\mathrm{gse}$), the number of galaxies that make up the GSE-event~($N_{\rm gse}$), and the stellar mass ratio of the progenitor's peak mass with the MW at the time of the merger~($M_\mathrm{gse}/M_\mathrm{MW}$).
Appendix~\ref{app:nehod} provides additional details regarding the training procedure and validation tests for this emulator.

As a first step, we consider the more general case of a Radially Anisotropic~(RA) event~\citep{2025Folsom}, which only satisfies the criteria on the stellar debris, $\beta>0.5$ and $f>0.5$.  
The left panel of Figure~\ref{fig:gse_prob} shows how the varied simulation parameters impact the fraction of generated MW-mass systems with an RA event.  
For each simulation parameter, we sample the parameter space uniformly, keeping the other four parameters fixed to their fiducial value, and generate $10^5$ samples consistent with the properties of the DREAMS galaxies with resolved disks. 
Binning by the specific simulation parameter, we count the fraction of all emulated galaxies that satisfy the RA condition.  
The 1$\sigma$ scatter, shown as the shaded band, contains uncertainty on the host total mass, intrinsic halo-to-halo variance, and epistemic uncertainty from the emulator training (see Appendix~\ref{app:nehod}).
As shown in Figure~\ref{fig:gse_prob}~(left), the SN energy parameter, $\widebar{e}_w$, has the most significant effect on the fraction of systems with an RA event. As \ew\,increases from the minimum to maximum of its range, the fraction of systems with an RA event decreases from $(27\pm1)$\% to $(14\pm2)$\%.  The other four parameters ($\Omega_{\rm m}$, $\sigma_8$, $\kappa_w$, $\epsilon_{f,{\rm high}}$) do not strongly affect this fraction. 

Using the emulated sample described above, we also implement the remaining selection criteria on merger mass ratio, accretion time, and quiescent history to get a sample of GSE-like events.  The results are shown in the right panel of Figure~\ref{fig:gse_prob}.  The overall fraction of systems with a GSE-like event goes down by nearly an order-of-magnitude from the RA case.  Again, the SN wind energy has the most significant effect, with the fraction of a GSE-like event decreasing from $(2.2\pm0.3)$\% to $(0.5\pm0.2$)\% moving from the minimum to the maximum of that range.
Overall, none of the simulation parameters affect the likelihood of a GSE analog merger by more than $\pm1$\% from 1.5\%, the expected result for the fiducial TNG parameters.

To understand the steep decrease in GSE-like events compared to RA events, we directly examine the simulations in the DREAMS CDM suite. Of the MW-mass hosts that have RA events, only 12\% have not had a major merger, defined as a stellar mass ratio of at most 1\,:\,50, in the last 8~Gyr.  This means that most of these galaxies would have had a more active merger history than expected for the MW, which appears to have quiet merger activity after the GSE event. Therefore, selecting the time and mass ratio of the GSE analog produces a much smaller sample of hosts than the RA criteria alone. 

The SN energy, \ew, has a strong effect on the fraction of GSE mergers because it is correlated with the infalling stellar material that ends up in the halo.  For the lowest~(highest) \ew, MW-hosts with GSE-like events have a mean $f\sim 0.48$~(0.29).
Especially strong SN winds are effective at ejecting gas from galaxies, which can slow the growth of the MW-mass progenitor, as well as its infalling satellites~(Garcia~et~al.,~in~prep).
This is also true for the speed of SN winds, where increasing \kw also lowers the mass of these satellites.  However, due to the minimum wind speed imposed, galaxies with low masses and at high redshift are not affected by variations to \kw.
Given that \kw\,does not seem to affect the likelihood of a GSE-event, it suggests that either the wind speed is set to $v_{\rm min}$ at these redshifts and masses, or that the wind velocity is less critical than the total energy injection for determining merger rates and debris fractions.

Our finding that GSE analogs are intrinsically rare is consistent with other studies using large-volume simulations like EAGLE~\citep{2020Evans} and TNG50~\citep{2025Folsom}, as well as other zoom-in campaigns like Auriga~\citep{2019Fattahi} and MW-est~\citep{2024Buch}.
\cite{2020Evans} presents a similar fraction of MW-mass systems having a GSE analog, at 5\%, but uses a strict stellar mass cut of $\mstar=\left(0.5\text{--}1\right)\times10^9~\mdot$ instead of the merger ratio used here.
\cite{2024Buch} requires a total mass ratio greater than 1\,:\,5, but does not require a radial merger.
They find that 41\% of the MW hosts with LMCs also have a GSE merger.
\cite{2019Fattahi} and~\cite{2025Folsom} only select RA mergers, not the full merger history as we do here, finding $\sim$30\% of their systems had RA mergers, although they use different limits on $\beta$.
The results presented in this subsection highlight that the fraction of GSE mergers predicted is highly sensitive to the criteria used to select them, especially when requirements are placed on the mass and accretion time of the original progenitor.


\subsection{Properties of MW-mass Galaxies with GSE Analogs}
\label{sec:gse_props}

\begin{figure*}
    \centering
    \includegraphics[width=\textwidth]{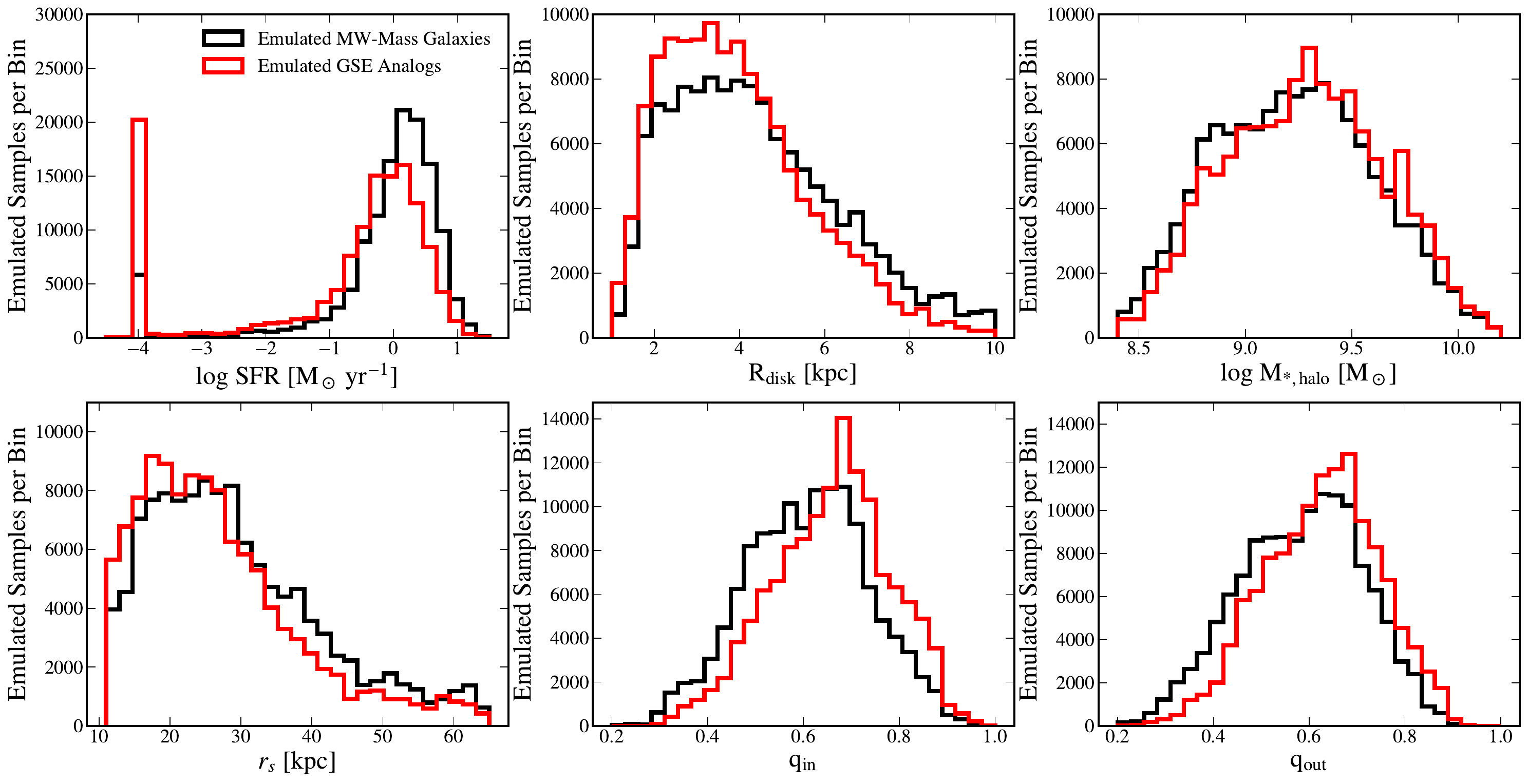}
    \caption{A comparison of key structural properties between the general emulated MW-mass galaxy population (black) and the subset of galaxies with GSE analogs (red).
    The six panels show distributions for the 
    star formation rate (SFR; top left),
    disk radial scale length ($R_\mathrm{disk}$; top middle), 
    stellar halo mass ($M_{*, \mathrm{halo}}$; top right), 
    stellar halo break radius ($r_s$; bottom left), 
    inner halo shape ($q_\mathrm{in}$; bottom middle,
    and outer halo shape ($q_\mathrm{out}$; bottom right).
    The population of galaxies with a GSE analog shows systematically different distributions for most properties, including a strong tendency toward lower SFRs, smaller $R_\mathrm{disk}$, and higher $q_\mathrm{in}/q_\mathrm{out}$.
    However, we find no significant systematic difference in $M_{*, \mathrm{halo}}$ between the two populations.
    }
    \label{fig:gse_props}
\end{figure*}

Given the established rarity of these merger histories, we now investigate their impact on the MW-mass host galaxy. Emulator training is performed according to the method outlined in Section~\ref{sec:emulator}, but with updates to the target variables.  A detailed discussion of the validation for this emulator is also given in Appendix~\ref{app:nehod}.  

For this dataset, the target variables include the five parameters related to the selection of GSE analogs ($t_{\rm mm}$, $t_{\rm gse}$, $M_{\rm MW}/M_{\rm gse}$, $f$, and $\beta$) and the 18 MW properties (those shown in Figure~\ref{fig:violin} and Table~\ref{tab:props}).
We only train on the 803 galaxies that have resolved disks, see Section~\ref{sec:props}.
Similar to the other trainings, we generate $2~\times10^5$ sample batches with the five trials with the lowest validation loss to account for the model uncertainty.
For these samples, all conditioning parameters ($M_\mathrm{TOT}$, \om, \s8, \ew, \kw, \agn) are varied uniformly through their entire parameter space.
We select the GSE sample as all galaxies from these $10^6$ samples that satisfy the five criteria outlined in Section~\ref{sec:gse_merger}.
This results in a sample of 13,613 samples with GSE analogs.

An expanded discussion of the GSE effects on MW hosts is provided in Appendix~\ref{app:props}.  There, the reader can find a corresponding  plot to Figure~\ref{fig:violin} showing how the host distributions change in the presence of a GSE-like event (Figure~\ref{fig:violin_gse}).  Figure~\ref{fig:1D_props} provides the distribution of all 18 MW properties studied here for the emulated sample with and without a GSE-like event, while Figure~\ref{fig:nehod_corner_gse} provides the correlations for a subset.  Figure~\ref{fig:nehod_corner_tng} shows how the results change when the astrophysical parameters are fixed to their TNG fiducial values.


Figure~\ref{fig:gse_props} shows a subset of the plots from Figure~\ref{fig:1D_props}, focusing on the following six MW properties: SFR, $R_\mathrm{disk}$, $M_{*,\mathrm{halo}}$, $r_s$, $q_{\rm in}$, and $q_{\rm out}$. As seen in the top-left panel, the most substantial shift between the population of general MW-mass hosts and those with a GSE analog is in the SFR, especially the fraction of host galaxies that are quenched at present day.  In particular, a host with a GSE analog is $2.7\times$ more likely to be quenched at $z=0$. Additionally, star-forming hosts are more likely to have lower SFR at present day if they have experienced a GSE-like event: the median log SFR decreases from $0.11$ to $-0.24$ $\text{log}\left(\mdot~\text{yr}^{-1}\right)$.
As a potential explanation for this trend, a gas-rich GSE progenitor merges with the MW radially at early times, it delivers gas to the central region of the host that is either accreted onto the BH or forms stars, producing a burst of star formation and enhanced AGN activity.
The feedback from the star formation and BH accretion could then expel a significant portion of the remaining gas from the galaxy, reducing its overall baryon content and lowering its SFR at late times.

While the mass of the host disk is not significantly affected, its radial scale length shifts to lower values in cases with a GSE analog. 
This shift brings the population of hosts with a GSE analog into better agreement with observations, increasing the percent of galaxies within the observational uncertainty range from 19\% to 25\%.
The circular velocity at 8~kpc, $V_\mathrm{circ}$, is also shifted toward higher values, likely due to the more compact disk and more massive bulge (see Figure~\ref{fig:1D_props}).

Surprisingly, the MW-mass galaxies do not have systematically different stellar halo masses at present day if they experienced a GSE-like event. This suggests that stellar halos similar to what we observe for our Galaxy can be created by a combination of accretion from minor mergers, contributions from $\emph{in-situ}$ stars kicked up from the disk, and/or other early major mergers.  Therefore, while the GSE event is rare and significant in the MW's history, it is not sufficient to explain the MW's exceptionally low stellar halo mass.
That being said, it can lead to differences in the structure of the stellar halo.  
The largest relative difference appears in the shape of the inner stellar halo, $q_\mathrm{in}$, which is more spherical in the population with GSE analogs---the median of the distribution shifts from 0.61 to 0.66 without and with a GSE, respectively.
Additionally, there is a systematic, but less significant, shift in the outer halo shape, $q_\mathrm{out}$, from 0.58 to 0.63.  
While these changes do not significantly alter the percentage of emulated galaxies that overlap with the observational range for most halo properties, they do increase the percentage of galaxies with a GSE analog and outer halo shapes similar to the MW from 23\% to 37\%.

Our finding that the GSE merger does not necessarily result in a systematically more massive stellar halo aligns with the recent conclusions of~\cite{2025Proctor}, who find a similarly weak connection in the EAGLE simulations.
They demonstrate that a low stellar halo mass does not correlate with a less active merger history and that $\sim$$25$\% of their disk galaxies undergo a merger with a satellite whose mass is at least 10\% of the host after $z=1$.
Our results complement this view by showing that even a significant merger like the GSE does not dictate the stellar halo mass at $z=0$.
Together, these results challenge the simple assumption that the stellar halo mass is a reliable tracer of a galaxy's past merger history.

Furthermore, both studies suggest that the merger history can significantly affect components other than the stellar halo.
\cite{2025Proctor} find that mergers on circular, co-planar orbits can build up a thick, extended stellar disk without affecting the stellar halo.
This process can rejuvenate the disk with fresh gas and lead to a burst of star formation.
In contrast, the DREAMS simulations show that MW-mass galaxies with a GSE analog tend to have more radially compact stellar disks. 
These different outcomes are likely due to the distinct orbital properties being studied.
While the co-planar mergers in~\cite{2025Proctor} directly contribute to the extended disk, the radial nature of the GSE merger combined with a subsequent long quiet period allows for the disk to regrow and become more compact over time.



These results underscore that while matching the GSE merger reproduces some of the MW's structural properties, it is not a complete explanation for its unique structure.
Likely, there are still other stochastic and environmental properties of the MW that are important to match to obtain a true MW analog.  Indeed, since the DREAMS simulation suite implements an environment selection criterion that excludes potential halo-mass-matched targets in dense regions, our analogs are systematically biased toward low-density environments.

\section{Conclusions}
\label{sec:conclusion}

This paper introduced a new hydrodynamical simulation suite that consists of 1,024 zoom-in MW-mass galaxies in a $\Lambda$CDM cosmology as part of the DREAMS Project.  Each simulated galaxy has a unique initial condition and a unique set of varied astrophysical and cosmological parameters.  This suite also comes with an N-body counterpart that can be used to study the effect of baryons on observables of interest.  Taken together, these simulations provide an unprecedented opportunity to characterize the halo-to-halo variance and theoretical modeling uncertainties on predictions for MW-mass systems, at least within the TNG model. 

Of the varied parameters, the three most impactful are related to the SN energy (\ew) and wind speed (\kw), as well as the efficiency of AGN feedback (\agn).  We presented a novel weighting scheme that assigns a weight to combinations of \ew-\kw-\agn\, that depends on how consistent the galaxies generated with those parameters are with the observed present-day SMHM relation.  We found large, degenerate regions of high-weight parameter space, suggesting that there is no unique parameter combination required to reproduce the observed SMHM.  The original TNG model is consistent with these results, lying within the extended region of high-weight parameter space recovered here, but is clearly not a unique solution.  These results demonstrate that a wide range of physical models can produce realistic galaxy populations and highlight the importance of exploring parameter uncertainty rather than relying on a single tuned model. The weighting procedure introduced here can be extended to other redshift-dependent and mass-varied scaling relations, which may further reduce the degeneracies on the input parameters.

Using the hydrodynamical DREAMS CDM suite, we characterized structural properties of the simulated MWs and compared them to our own Galaxy.
The main results are:

\begin{itemize}
    \item It produces a diverse population of MW-mass galaxies whose properties encompass the observed properties of the MW, including its rarer features like an undermassive SMBH and low-mass stellar halo.
    \item For all galactic properties that we measure, except for $M_\mathrm{BH}$, the scatter is dominated by a combination of MW halo mass uncertainty and intrinsic halo-to-halo variance, demonstrating the need for statistical samples of galaxy populations to disentangle the impact of events and characteristics that are specific to the MW's history.
    \item A MW-like merger history, with an early GSE-like merger and no subsequent major mergers, is rare, occurring in only 1.5\% of the 803 disk (or 1.2\% of the total 1,024) galaxies in the suite.
    \item Galaxies that include a GSE analog in their merger history systematically differ from the general population, exhibiting lower SFRs, more compact disks, and more spherical stellar halos. However, significant scatter remains for all properties due to halo-to-halo variance.
\end{itemize}


Beyond direct comparison, the DREAMS approach offers distinct methodological advantages.
Other approaches, like using genetic modification to create constrained simulations of the MW's environment and formation history~\citep{2018Rey, 2021Stopyra}, are limited in that they are only applicable to the chosen environment and often adopt a single galaxy-formation model~\citep{2021Davies,2021Orkney,2023Rey}.
DREAMS offers a complementary approach through emulation, which can generate statistical samples of galaxies with rare properties, like GSE analog mergers, while marginalizing over physics uncertainties.
Additionally, the DREAMS approach enables the reverse inference process where $z=0$ properties can be used to reconstruct a galaxy's merger history~\citep{2023Eisert, Leisher2025}.
The DREAMS suite, with its built-in model uncertainty and large sample of galaxies, provides an ideal training set for such methods that could be used to reveal additional details about the MW's history.

The MWs in this suite are relatively isolated, with no requirement of an LMC- or M31-like neighbor.  This may affect the predictions of some of the structural properties that are compared to observations.  For example, requiring an LMC-like merger would likely shift the distribution for total stellar mass, $M_*$, and satellite stellar mass, $M_{*, {\rm sat}}$, into better alignment with observed values.  

There are three key ways that the DREAMS CDM suite can be expanded in the future.  The first is to increase the number of parameters varied within the TNG model, similar to what was done by~\cite{SB28} for CAMELS. This would improve quantification of uncertainties within a given sub-grid framework.  
The second is to expand to additional galaxy-formation models, which would quantify uncertainties between sub-grid frameworks.
The TNG model relies on its subgrid model for many of the subresolution physics that cannot be directly modeled~\citep{2017Weinberger, 2018Pillepicha}.
For example, SMBH growth is based on a subgrid prescription tied to the local gas density, but does not directly resolve the multiphase ISM or accretion disk~\citep{Sijacki2015, Bhowmick2022}.
Expanding to models with different multiphase ISM or accretion disk models would help to marginalize over these additional theoretical uncertainties.
The third is by altering the selection criteria for the simulated MWs, requiring an LMC or M31 companion at present-day or a previous GSE-like merger.



This paper will be accompanied by three additional papers studying the implications of the new DREAMS CDM suite to the MW's (1)~satellite population~(Rose~et~al., in~prep), (2)~DM density distribution~(Garcia~et~al., in~prep), and (3)~DM velocity distribution in the solar neighborhood~(Lilie~et~al., in~prep).  Taken together, these works underscore the fundamental challenge of galactic archaeology: disentangling the interwoven effects of a galaxy's unique formation history from uncertainties in the underlying physical models.
A DREAMS-like approach, which combines systematic variations over initial conditions and model parameters, is essential for this task.


\section*{Acknowledgements}

We thank the Simons Foundation for their support in hosting and organizing workshops on the DREAMS Project.
Additionally, we gratefully acknowledge the use of computational resources and support provided by the Scientific Computing Core at the Flatiron Institute, a division of the Simons Foundation.
ML acknowledges support from the Simons Investigator Award.  AMG, NK, PT, and AF acknowledge support from the National Science Foundation under Cooperative Agreement 2421782 and the Simons Foundation grant MPS-AI-00010515 awarded to NSF-Simons AI Institute for Cosmic Origins -- CosmicAI, \href{https://www.cosmicai.org/}{https://www.cosmicai.org/}. The work reported on in this paper was performed, in part, using Princeton University's Research Computing resources.

\section*{Data Availability}

The simulation data presented in this paper are publicly available through the DREAMS website at \url{https://www.dreams-project.org}. The code used to perform the analysis is available in the DREAMS repository at \url{https://github.com/DREAMS-Project} and the NeHOD repository at \url{https://github.com/trivnguyen/nehod_torch}.

\appendix

\section{Emulator Validation}
\label{app:nehod}

\setcounter{equation}{0}
\setcounter{figure}{0} 
\setcounter{table}{0}
\renewcommand{\theequation}{A\arabic{equation}}
\renewcommand{\thefigure}{A\arabic{figure}}
\renewcommand{\thetable}{A\arabic{table}}

This appendix describes the architecture for the conditional normalizing flow emulators used in this work (Appendix~\ref{sec:architecture}), as well as a number of validation tests (Appendices~\ref{app:reproduction}--\ref{app:extrapolation}).

\subsection{Emulator Architecture}
\label{sec:architecture}

The emulators used in this work are based on the neural spline flow architecture that makes up the first component of the architecture used in~\cite{NeHOD}.
They are essential for efficiently exploring the five-dimensional simulation parameter space and generating the large statistical sample of galaxies required for our analysis.

We employ one core architecture but conduct two distinct trainings to produce emulators tailored for different scientific goals.
The first emulator focuses on the parameter weighting scheme (Section~\ref{sec:weights}) and the second on the merger history analysis (Section~\ref{sec:gse}).
The former is conditioned on five simulation parameters (\om, \s8, \ew, \kw, and \agn) and trained to predict the $\mstar$ and $\mtot$ required for the weighting scheme in Section~\ref{sec:weights}.
The second emulator is conditioned on the same five input parameters plus $\mtot$ and trained to predict parameters used in the GSE selection (e.g., $t_{\rm mm}$, $f$, $\beta$) and properties of the Milky Way~(MW)-mass host (e.g., SFR, $M_\mathrm{BH}$, $V_{\rm circ}$).
Each emulator consists of a sequence of transformations where each is parameterized by a neural network with four hidden layers and 32 units.

The models are trained to minimize the negative log likelihood of the target properties given the conditioning parameters.
We use the \textsc{AdamW} optimizer~\citep{adamW} with a batch size of 32. The learning rate is managed by a scheduler that includes a linear warm-up phase of 100 steps followed by a cosine annealing decay.
The training is set for a maximum of $5\times10^4$ steps.
To prevent overfitting, we monitor the loss on a held-out validation set consisting of 20\% of the simulations and implement early stopping with a patience of 500 steps. For more details on the architecture specifics, see~\cite{NeHOD}.

To optimize the model performance, we conduct two hyperparameter searches (one per emulator), each using \textsc{optuna}~\citep{optuna}, with the weights emulator undergoing 100 trials and the merger history emulator undergoing 50.
We tuned the learning rate ($10^{-5}\text{--}10^{-1}$), weight decay ($10^{-4}\text{--}0.5)$, dropout rate ($10^{-4}\text{--}0.5$), and flow depth (1--10 transforms).
For the weights emulator,  the hyperparameter optimization identified that the best-performing model configuration has a learning rate of $2.6\times10^{-3}$, weight decay of 0.21, dropout rate of $7.7\times10^{-2}$, and a flow depth of 1 transformer.
For the GSE emulator, the hyperparameter optimization identified that  the best-performing model configuration has a learning rate of $8.1\times10^{-4}$, weight decay of $5.8\times10^{-2}$, dropout rate of 0.04, and a flow depth of 8 transformers.

To incorporate uncertainty in the emulation process itself, we construct an ensemble of the best-performing models (5 for the GSE emulator and 10 for the weights emulator)---those with the lowest validation loss---from the hyperparameter searches with \textsc{optuna}.
The final generated datasets are the combined outputs from these best-performing models.

\subsection{Distribution Reproduction}
\label{app:reproduction}

\begin{figure*} 
    \centering
    \includegraphics[width=\textwidth]{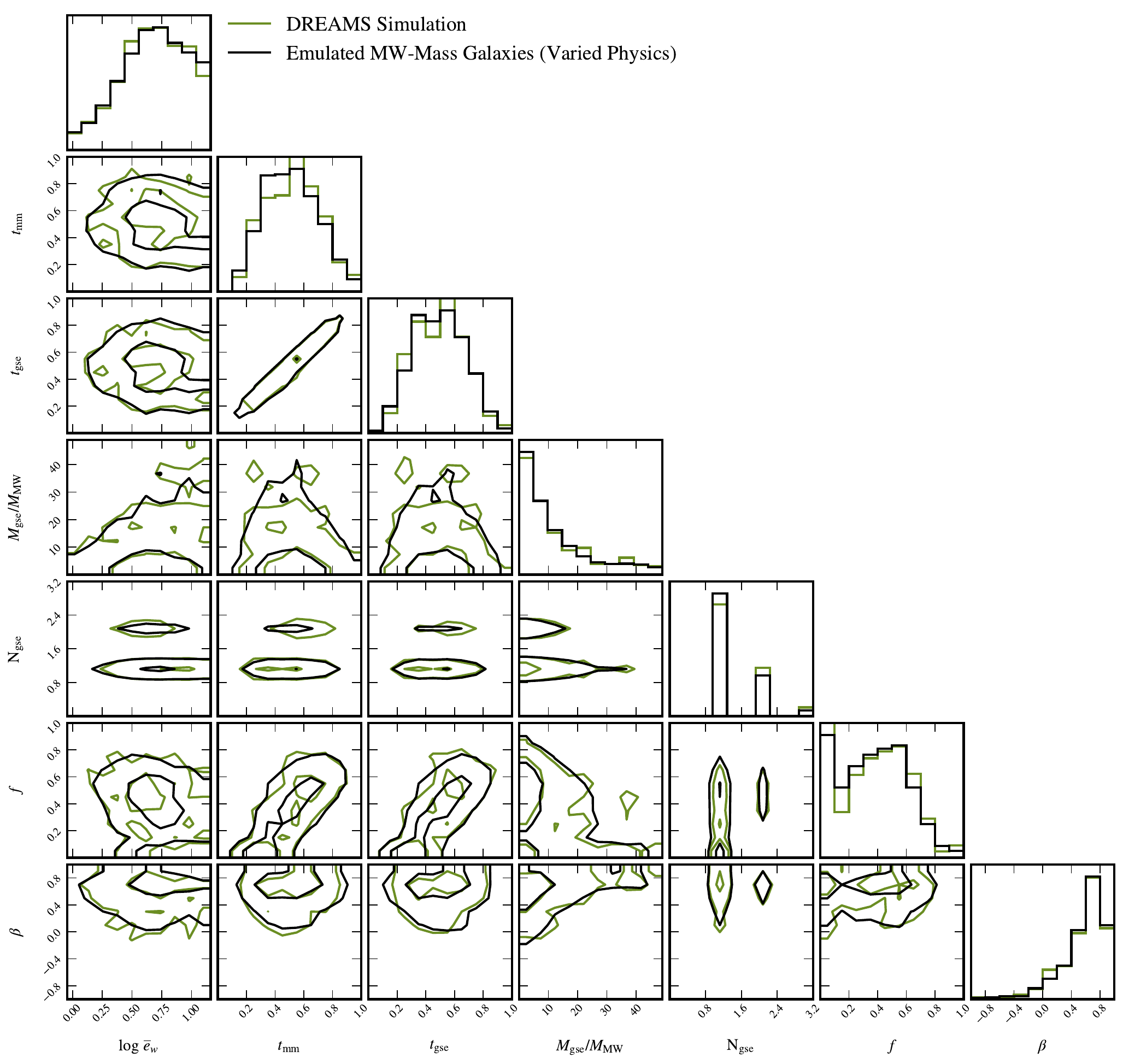}
    \caption{Validation of the emulator used in Section~\ref{sec:gse}.
    This corner plot compares the joint distributions of six properties that are used to select GSE analogs and the most influential feedback parameter.  The two-dimensional contours correspond to the 1 and 2$\sigma$ containment regions.  
    The predictions from the emulator~(black) are compared against the ground truth from the 803 disk galaxies in the DREAMS simulations~(green).
    The properties shown are the energy of SN feedback~(\ew), the scale factor of the last major merger~($t_{\rm mm}$), the scale factor of the GSE candidate merger~($t_{\rm gse}$), the stellar-mass merger ratio of the MW and GSE candidate~($M_{\rm MW}/M_{\rm gse}$), the number of galaxies that contribute to the GSE-like event~($N_{\rm gse}$), the fraction of $z=0$ stellar halo debris from the GSE-like event~($f$), and the radial anisotropy of that debris~($\beta$).
    The one-dimensional histograms along the diagonal and the two-dimensional contours in the off-diagonals show excellent agreement.
    }
    \label{fig:nehod_corner_gse_selection}
\end{figure*}

We validate the accuracy of the emulator sample by comparing the generated probability distributions against the ground truth distributions from the DREAMS CDM suite.
This step ensures that the emulator correctly captures both the intrinsic halo-to-halo variance of the galaxy population and the specific correlations between physical properties.

Figure~\ref{fig:nehod_corner_gse_selection} presents a corner plot for the emulator used in the GSE analysis (Section~\ref{sec:gse}; black lines) and the DREAMS simulations used to train and validate the model (green lines).
It compares joint distributions of the six properties used to identify the GSE analogs: the scale factor of the last major merger~($t_{\rm mm}$), the scale factor of the GSE-like merger~($t_{\rm gse}$), the stellar-mass merger ratio of the host and GSE-like merger~($M_{\rm MW}/M_{\rm gse}$), the number of galaxies that contribute to the GSE-like event~($N_{\rm gse}$), the fraction of $z=0$ stellar halo debris from the GSE-like event~($f$), and the radial anisotropy of that debris~($\beta$).
To show that the emulator is also capturing the dependence on the input parameters, we include the supernova~(SN) energy parameter, \ew.
The distribution for both the simulated and emulated samples are weighted according to the scheme presented in Section~\ref{sec:weights}.

The plot demonstrates excellent agreement between the emulated and true datasets, as seen in both the one-dimensional histograms and two-dimensional contours.
Along the diagonal, the emulated distributions successfully reproduce the shape, peak, and scatter of the simulation data for each individual property.
In the off-diagonal panels, the emulated contours closely track the true data, indicating that the model has learned the complex covariances between the galactic properties.
Crucially, the model reproduces the sharp boundaries and complex covariances inherent in the merger-history data, such as the correlation between $f$ and the two timing parameters, $t_{\rm mm}$ and $t_{\rm gse}$.

\begin{figure*} 
    \centering
    \includegraphics[width=\textwidth]{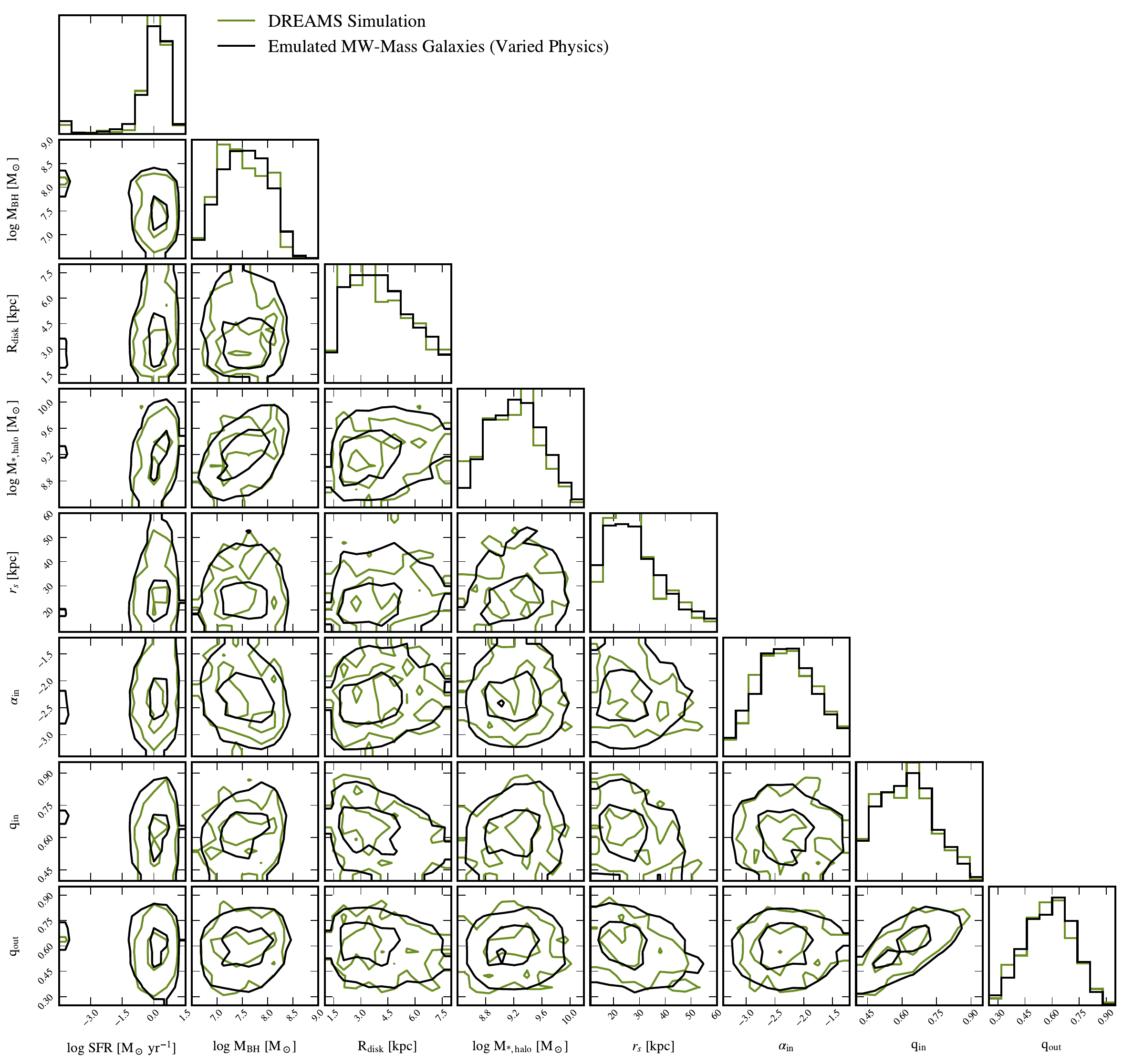}
    \caption{Validation of the emulator presented in Section~\ref{sec:gse}. 
    Similar to Figure~\ref{fig:nehod_corner_gse_selection}, this corner plot compares distributions generated by the emulator~(black) with the ground truth DREAMS simulations~(green).
    The plot shows a representative subset of the galactic properties that are most discussed in the text---see Section~\ref{sec:props} for definitions.
    The strong agreement confirms the emulator's ability to reproduce a wide range of galactic properties.
    }
    \label{fig:nehod_corner_MW}
\end{figure*}

Figure~\ref{fig:nehod_corner_MW} extends this validation to the general structural properties of the host galaxy.  For readability, only a subset of the properties are shown in the corner plot, but all 18 (see Table~\ref{tab:props}) have been checked.  As with the merger history properties, the emulator successfully recovers the diverse range of galactic structures found in the DREAMS suite.

These validation tests are critical for ensuring that the selection and analysis of GSE analogs work as intended.  
We also check that the full corner plots for all 29 parameters in the GSE emulator and the 7 parameters in the weights emulator are accurate, separately.
An inaccurate emulator could artificially favor or disfavor certain regions of parameter space, not because they produce unrealistic galaxies, but because the emulator itself is faulty.
Crucially, the emulator accurately reproduces the intrinsic scatter of the simulation properties, ensuring that the weighting scheme correctly penalizes models based on the full distribution.
The agreement between the emulated and simulated distributions ensures that the population statistics derived in the main text are robust representations of the underlying galaxy-formation model.


\subsection{Physical Dependencies}
\label{app:trends}

Next, we validate the accuracy of the emulator's conditional dependencies between the simulation parameters and host properties.
This ensures that the model has learned the underlying physics from the inputs rather than simply memorizing the training set.
In addition, these results demonstrate the physical trends that are the basis of the results presented in Sections~\ref{sec:weights}~and~\ref{sec:gse}.

\begin{figure*}[h]
    \centering
    \includegraphics[width=0.33\textwidth]{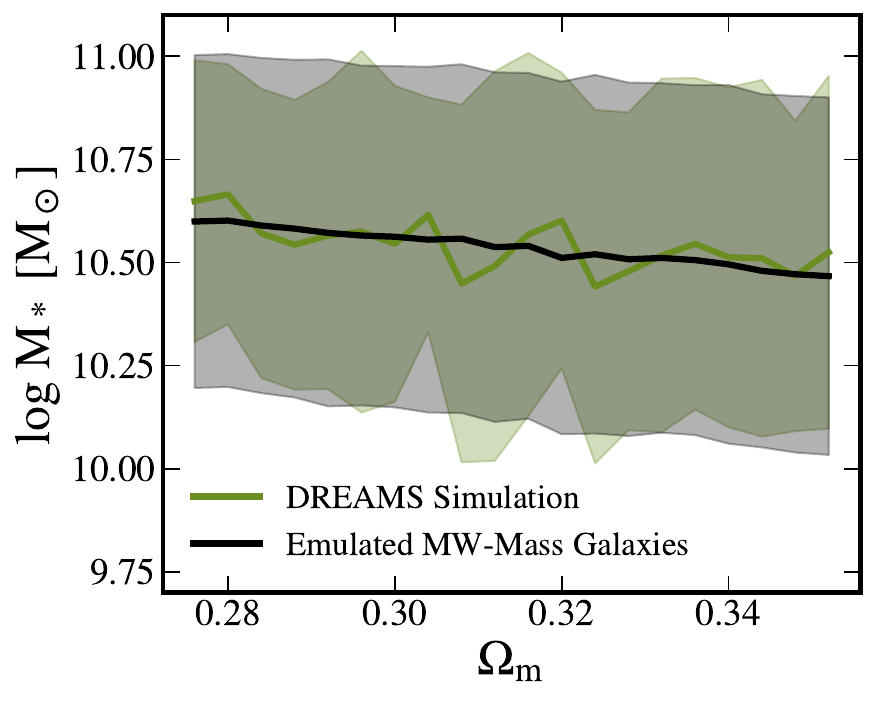}
    \includegraphics[width=0.33\textwidth]{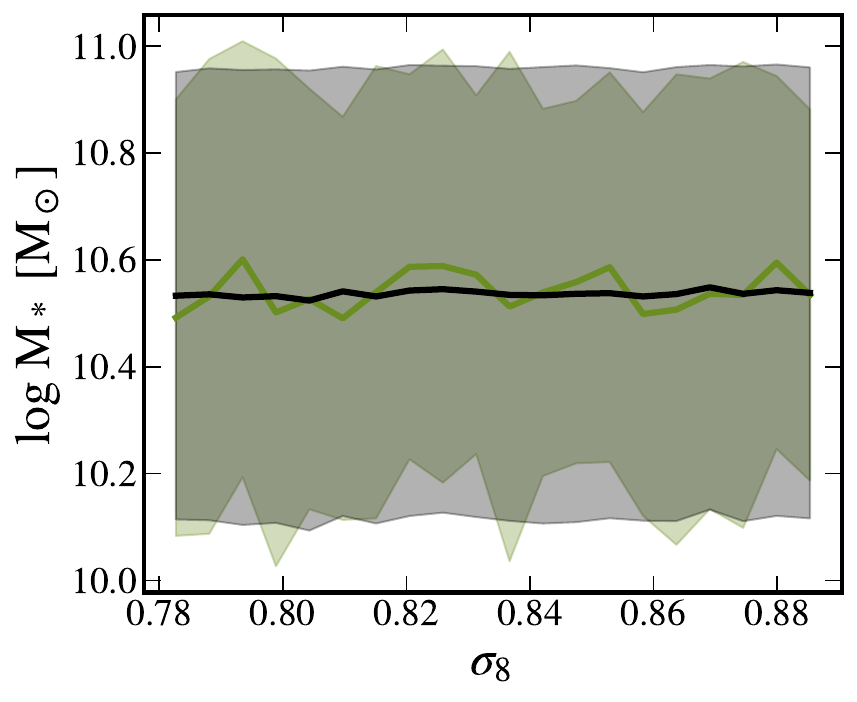}
    \includegraphics[width=0.33\textwidth]{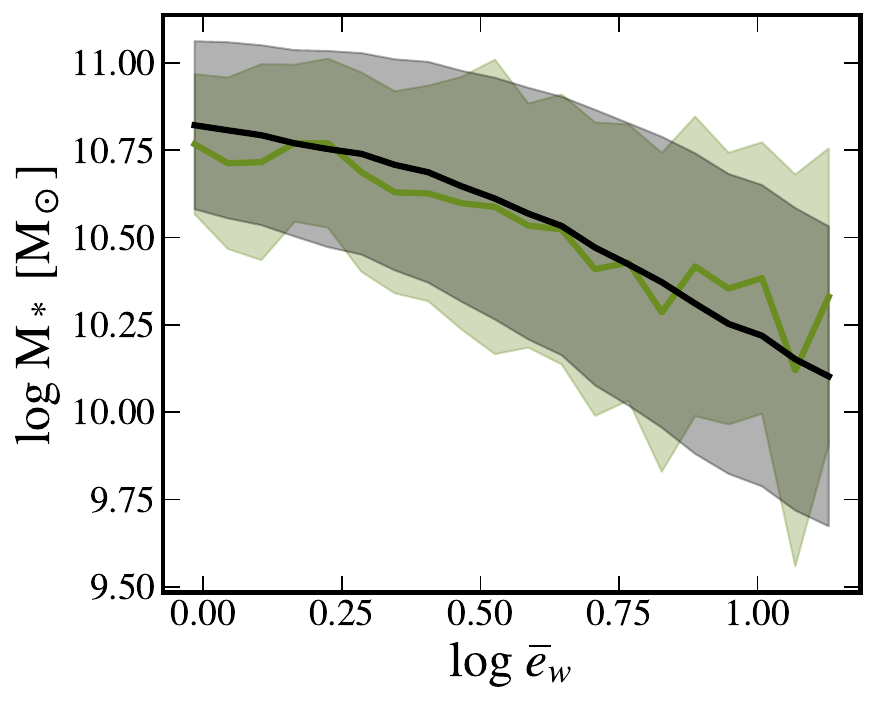}
    \includegraphics[width=0.33\textwidth]{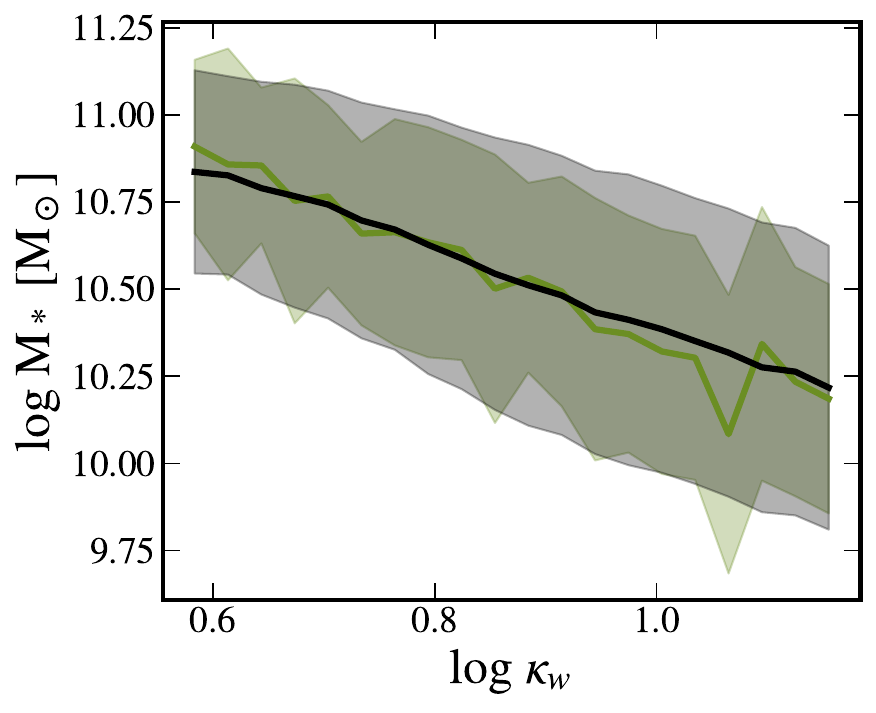}
    \includegraphics[width=0.33\textwidth]{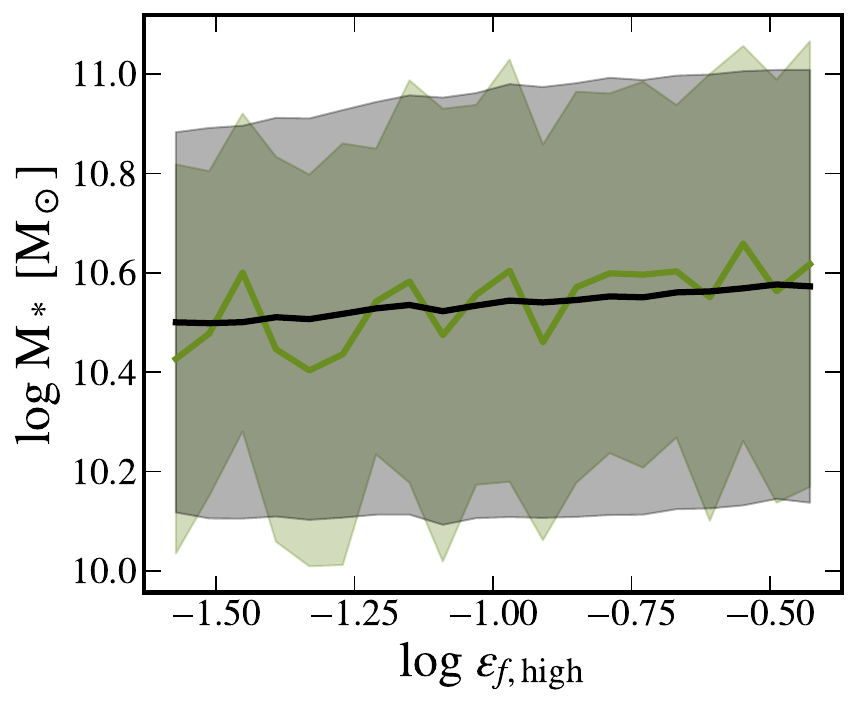}
    \caption{The sensitivity of the host galaxy's stellar mass to the five varied simulation parameters. 
    Solid lines represent the mean stellar mass in each bin, and the bands represent the $1\sigma$ uncertainty, for the DREAMS simulations~(green) and the emulated data~(black). 
    The emulator can correctly reproduce both the trends and the scatter of each parameter on the host's stellar mass.
    }
    \label{fig:Mstar}
\end{figure*}

Figure~\ref{fig:Mstar} examines the response of the host galaxy's stellar mass, $\mstar$, to variations in the five simulation parameters.
The solid lines represent the mean relation in each bin, while the shaded bands indicate the 1$\sigma$ scatter due to halo-to-halo variance, the other four input parameters, and uncertainty on the halo mass. The emulator~(black) accurately reproduces the physical trends present in the DREAMS simulations~(green).
As expected from the TNG model, increasing SN feedback energy~(\ew) or wind velocity~(\kw) suppresses star formation, leading to a monotonic decrease in stellar mass~\citep{2018Pillepicha}.
The other parameters, \om, \s8, and \agn, have negligible impact on $\mstar$ at this mass scale, when marginalizing over all other parameters.
The emulator correctly predicts both the trend and magnitude of the scatter at each point in the parameter space.
This is essential for the weighting scheme described in Section~\ref{sec:weights} to function as a proper likelihood evaluation.

\begin{figure*}[h]
    \centering
    \includegraphics[width=0.33\textwidth]{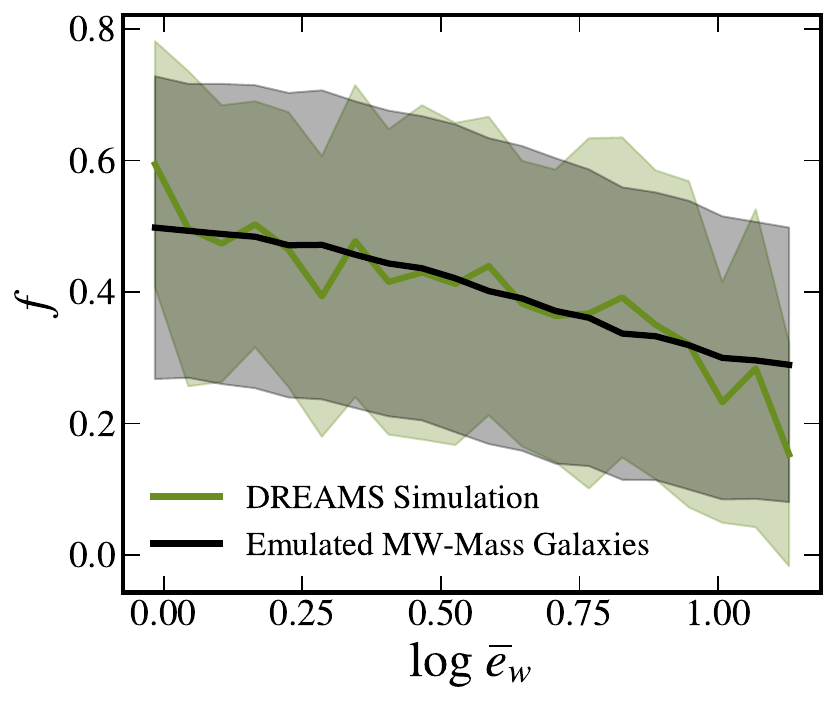}
    \includegraphics[width=0.33\textwidth]{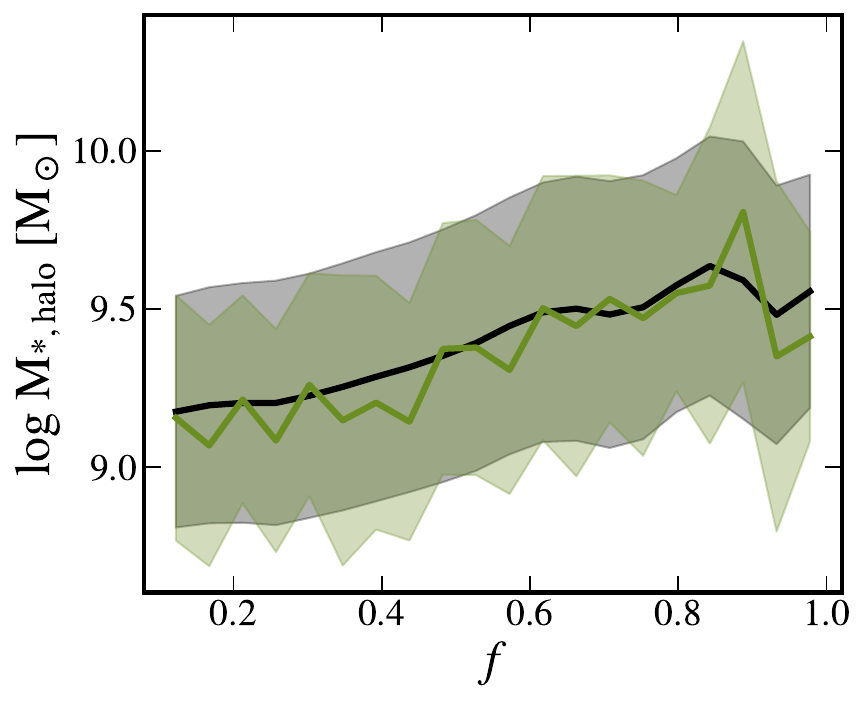}
    \includegraphics[width=0.33\textwidth]{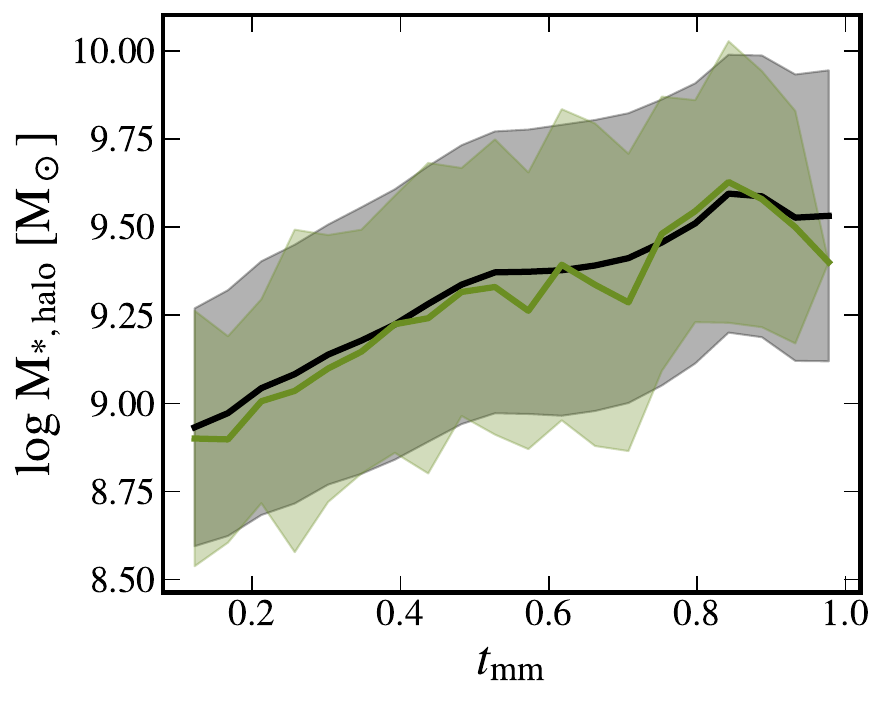}
    \includegraphics[width=0.33\textwidth]{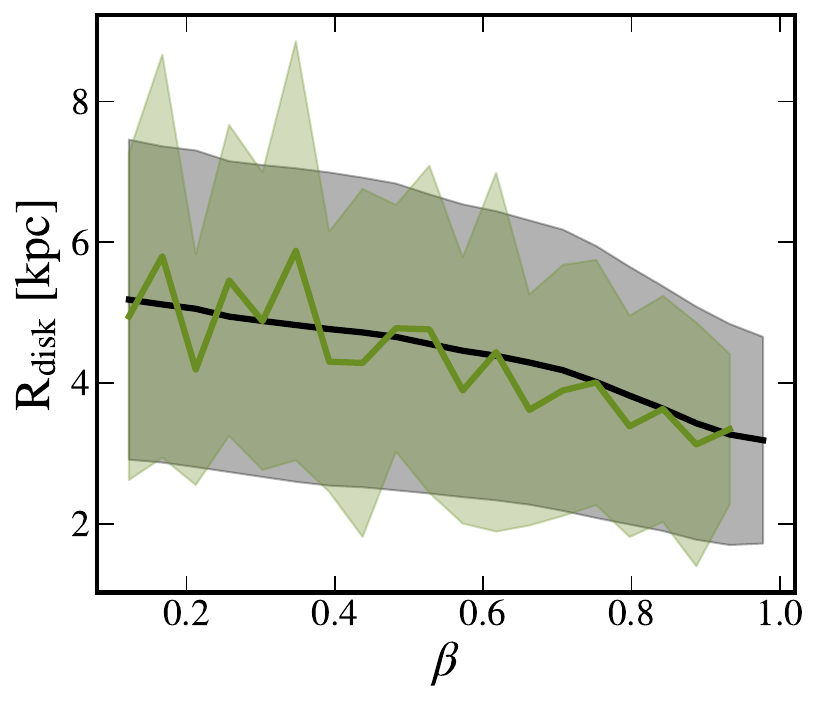}
    \includegraphics[width=0.33\textwidth]{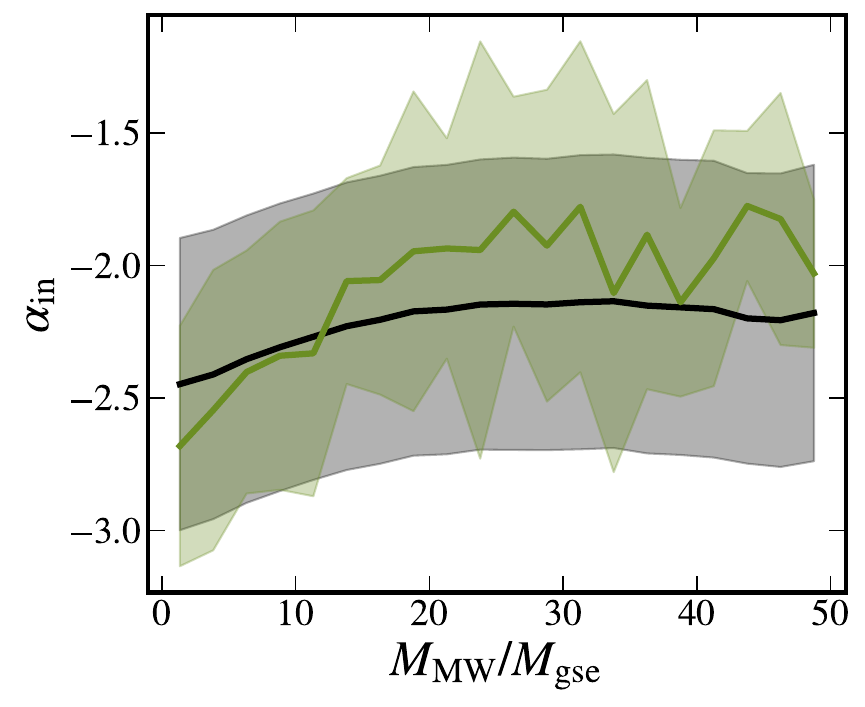}
    \includegraphics[width=0.33\textwidth]{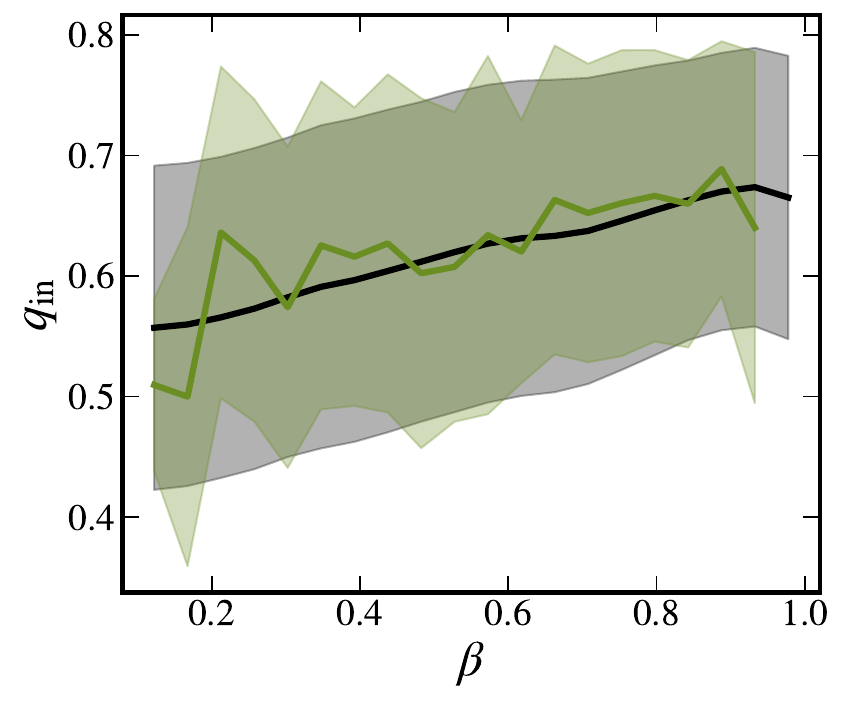}
    \caption{A subset of the correlations between key GSE and halo properties, as well as simulation parameters.
    The emulated MW-mass galaxies~(black) capture both the mean and $1\sigma$ in the DREAMS simulations~(green), giving confidence that the emulator is successfully learning the dependencies.
    While only a subset are shown here, we find that the emulator matches all property and parameter combinations well except for $M_{\rm MW}/M_{\rm gse}$ versus $\alpha_{\rm in}$ in the middle-bottom panel---see Appendix~\ref{app:nehod} for a discussion.
    }
    \label{fig:1D_trends}
\end{figure*}

Figure~\ref{fig:1D_trends} extends this test to a subset of the properties used in the GSE analysis.
We show the conditional dependencies of \ew, $f$, $\beta$, $t_{\rm mm}$, $R_{\rm disk}$, $M_{*,\mathrm{halo}}$, $\alpha_{\rm in}$, and $M_{\rm MW}/M_{\rm gse}$.
This subsample is chosen because the plots exhibit the strongest trends; most other parameter combinations result in no correlations.
For example, the dependence of $f$ on \ew\, is what drives the results shown in Figure~\ref{fig:gse_prob}. 
Additionally, the correlation between $\beta$ and $R_{\rm disk}$ or $q_{\rm in}$ shows how selecting a galaxy with a GSE analog (high $\beta$) can result in a more compact disk or spherical inner halo. 
It is important to note, however, that secondary trends may also be important in interpreting Figure~\ref{fig:gse_props}, which are only captured by the full 30-parameter latent space.

The only case where the emulator does not accurately capture the underlying distribution is the middle-bottom panel of 
Figure~\ref{fig:1D_trends}, which compares $M_{\rm MW}/M_{\rm gse}$ and $\alpha_{\rm in}$.  For this reason, we do not highlight the $\alpha_{\rm in}$ trend in the main text, but we do find that if the $M_{\rm MW}/M_{\rm gse}<5$ criterion is removed from the GSE selection, the discrepancy is reduced.

Overall, the emulator successfully captures the various relationships between the galactic parameters.  The agreement between the ground truth simulations and the emulated samples validates the use of the emulator for exploring physical drivers of the MW's assembly history.


\clearpage
\subsection{Convergence Tests}

Focusing on the weights emulator, this subsection assesses its convergence with respect to both the size of the training dataset and the number of generated samples.
These tests establish that our choice of 1,024 simulations is sufficient to capture the complex dependencies of the galaxy-formation model and that the generated datasets used to determine the weights are statistically stable.

\begin{figure*}[b]
    \centering
    \includegraphics[width=0.49\textwidth]{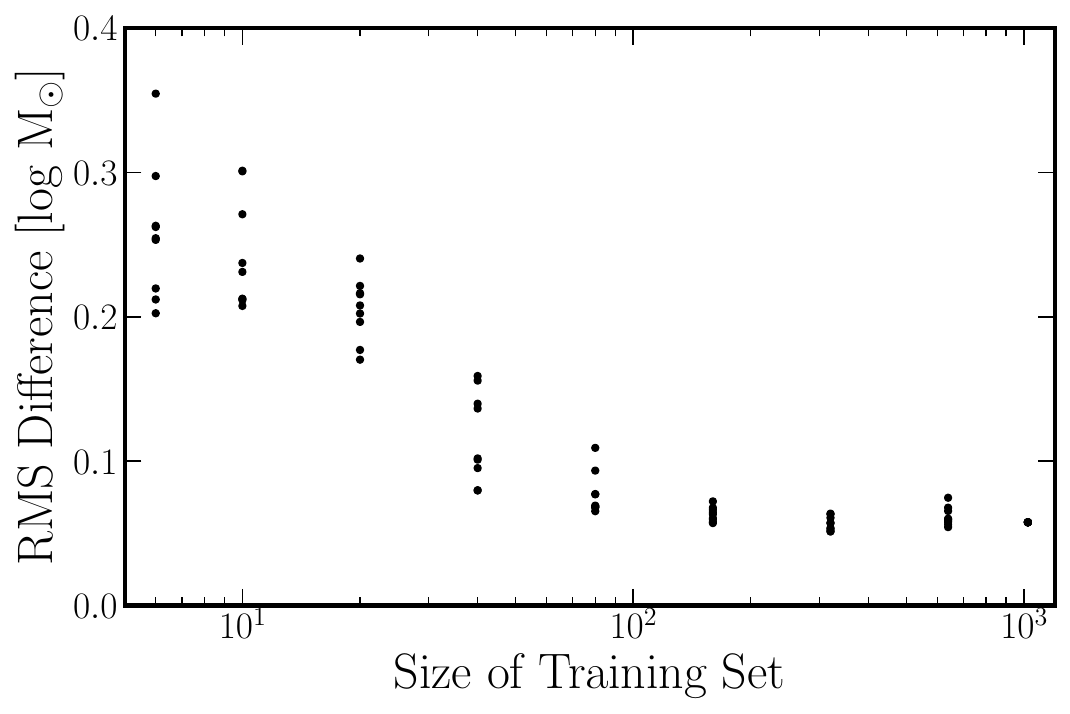}
    \includegraphics[width=0.49\textwidth]{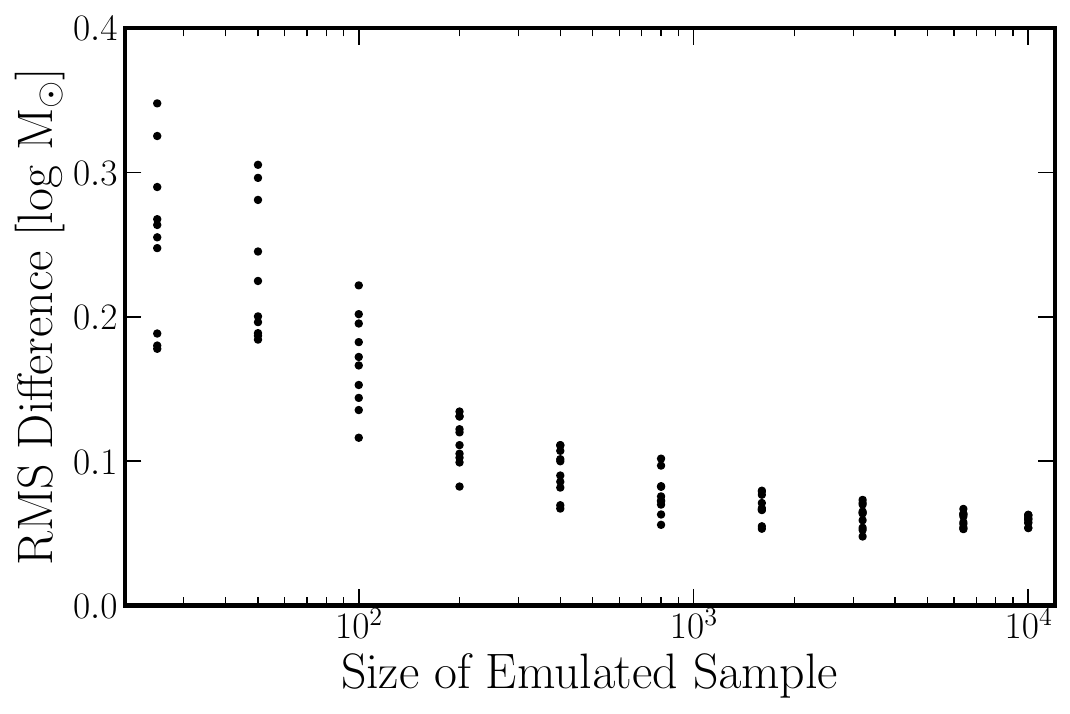}
    \caption{Convergence testing for the weights emulator.
    \textit{Left:} The RMS difference between the true and predicted SMHM relation as a function of the training-set size.
    \textit{Right:} The RMS difference as a function of the number of emulated samples generated.
    The multiple points plotted at each training set size and emulated sample size represent distinct emulators.
    Each emulator is trained using a different randomly selected subset of galaxies for the training data or a unique combination of hyperparameters for the emulated sample size data.
    The vertical scatter illustrates the variance in model performance driven by the specific choice of training examples and model architecture.
    The RMS difference plateaus above a training size of $\sim$200 simulations and an emulator sample size of $\sim$$10^3$, indicating that the full suite of 1,024 simulations are sufficient for calculating the parameter weights. 
    }
    \label{fig:sim_count}
\end{figure*}

The left panel of Figure~\ref{fig:sim_count} quantifies the accuracy of the emulator as a function of the training-set size.
We retrain separate emulators on randomly selected subsets of the DREAMS CDM suite, ranging in size from 6 to 1,024 simulations.
We then measure the root-mean-square~(RMS) difference between the predicted stellar mass--halo mass~(SMHM) relation and the true relation from all 1,024 DREAMS simulations.
The accuracy improves rapidly as the training set grows, with the RMS difference decreasing up to $\sim$200 simulations.
Beyond this point, the performance plateaus, indicating that the full suite of 1,024 simulations provides a well-converged basis for this analysis.

To robustly quantify the stability of our machine learning approach, we trained multiple independent emulators.
For every realization, we randomize either the specific subset of DREAMS simulations used for training (done only for the training set size test) or the hyperparameters governing the emulator's architecture (done only for the emulated sample size test).
By varying the galaxies used in the training set, we account for the impact of specific outliers or representative samples within small training sets.
By varying the hyperparameters, we marginalize over the epistemic uncertainty of the model configuration.
Consequently, the scatter observed at each column in Figure~\ref{fig:sim_count} reflects the variance from both the data selection and model tuning, demonstrating that our convergence is robust to these stochastic factors.

\begin{figure*}[t]
    \centering
    \includegraphics[width=\textwidth]{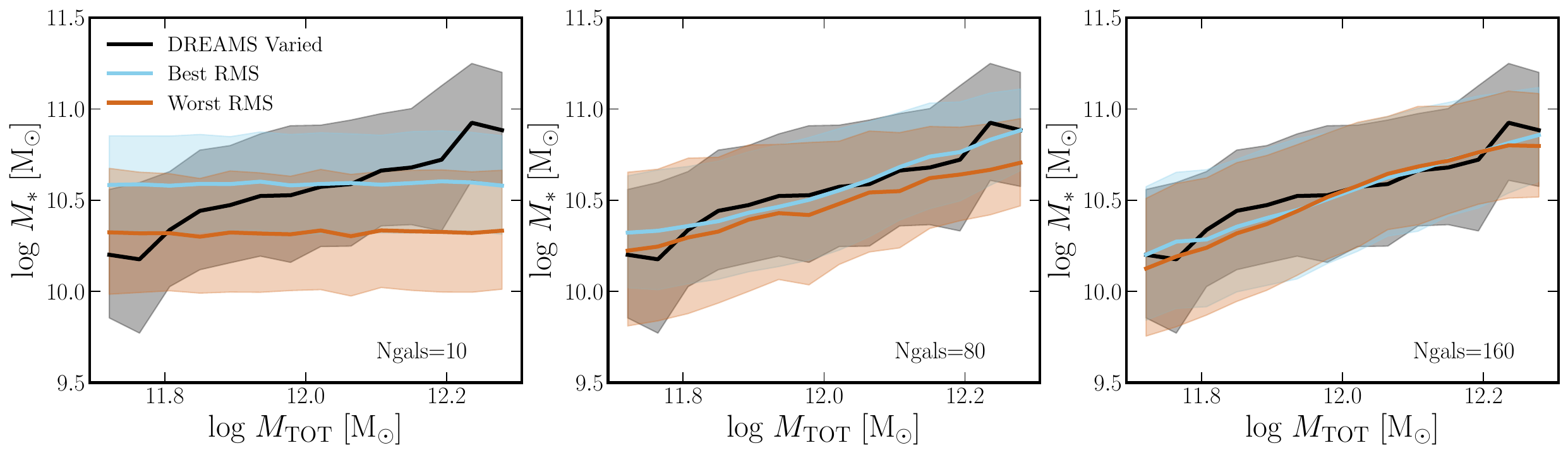}
    \caption{A visual representation of the RMS difference shown in Figure~\ref{fig:sim_count}.
    Each panel shows the mean and 1$\sigma$ spread of the emulated SMHM relation for ensembles generated using $\text{Ngals}=10, 90, \text{ and } 160$ simulations.
    We show two emulator results for each panel, the worst trained model (brown; the one with the highest RMS in Figure~\ref{fig:sim_count}) and the best trained model (light blue; the one with the lowest RMS in Figure~\ref{fig:sim_count}) for the given training set size.
    As the sample size increases, the emulator's prediction~(color bands) tightens and converges to the simulated result~(black band).
    }
    \label{fig:sim_count_visual}
\end{figure*}

Figure~\ref{fig:sim_count_visual} visually demonstrates this convergence, showing how the mean and predicted 1$\sigma$ scatter of the SMHM relation converges for the best-trained model (light blue) and worst-trained model~(brown), compared to the ground truth DREAMS simulations~(black). 
The best- and worst-trained models are chosen as those that produce the lowest and highest RMS error, respectively.
The panels correspond to training sets of $\text{Ngals} = 10, 80, \text{ and } 160$ simulations.  Even with only 160 simulations in the training set (resulting in a total of $\sim$200 simulations with the validation set), the worst-trained emulator predicts an accurate SMHM relation.

The right panel of Figure~\ref{fig:sim_count} examines the stability of the emulated population statistics with respect to the number of generated samples.
We measure the RMS fluctuations in the predicted SMHM relation while varying the emulator size from 10 to $10^4$ samples.
The RMS stabilizes once the sample size exceeds $\sim$$10^3$, confirming that the stochastic noise from the generation process is negligible for datasets used in our weights analysis.
Together, these tests demonstrate that the weighting emulator is not limited by the number of training simulations or the output sample size.
We note that we have not performed this same test for the full 29-parameter emulator in Section~\ref{sec:gse}.
This simple test does not ensure that its more complex dataset is fully converged across all 29 parameters; however, the agreement between the simulations and emulated samples examined throughout Appendix~\ref{app:nehod} gives us confidence that the emulator is operating within the regime of statistical stability for the given data.

\subsection{Interpolation versus Extrapolation}
\label{app:extrapolation}

\begin{figure*}[b]
    \centering
    \includegraphics[width=0.5\textwidth]{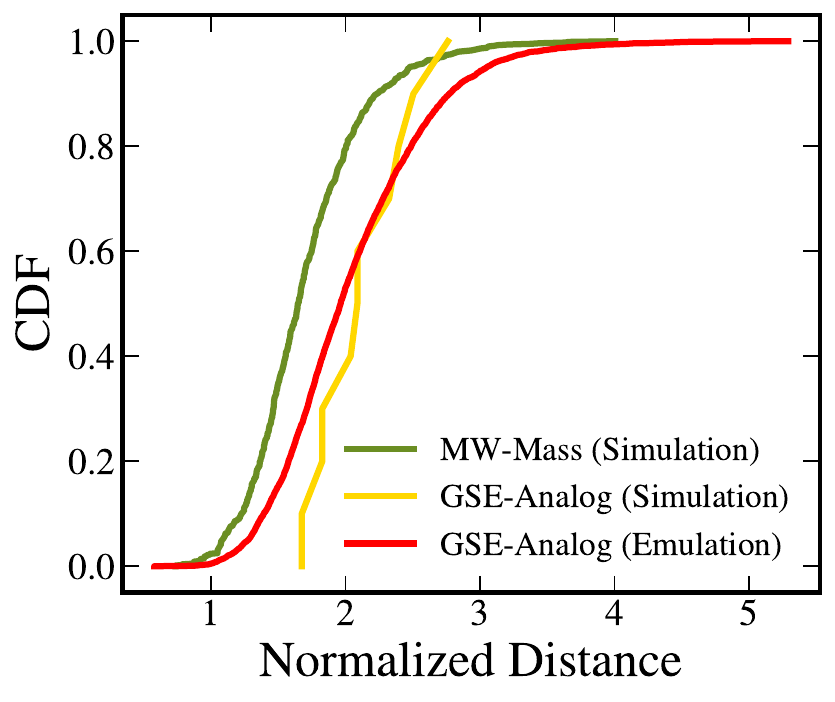}
    \caption{Cumulative Distribution Function (CDF) of the 13-dimensional normalized distances between samples in the latent space.
    The 13 parameters are key ingredients of the GSE analysis in Section~\ref{sec:gse}: $\widebar{e}_w$, $\kappa_w$, $t_{\rm mm}$, $t_{\rm gse}$, $M_{\rm MW}/M_{\rm gse}$, $f$, $\beta$, SFR, $R_{\rm disk}$, $M_{*,\mathrm{halo}}$, $r_s$, $\alpha_{\rm in}$, and $q_{\rm in}$.
    The green line shows the intrinsic distance for simulations in the DREAMS suite.
    The gold line shows the distance between each simulated GSE analog and its nearest simulated GSE analog neighbor.
    The red line shows the distance between the emulated GSE analogs and their nearest simulated GSE analog.
    The overlap between the red and gold lines indicates that the emulator is interpolating within the supported region of parameter space rather than overfitting or hallucinating outliers.
    }
    \label{fig:cdf}
\end{figure*}

\begin{figure*}
    \centering
    \includegraphics[width=\textwidth]{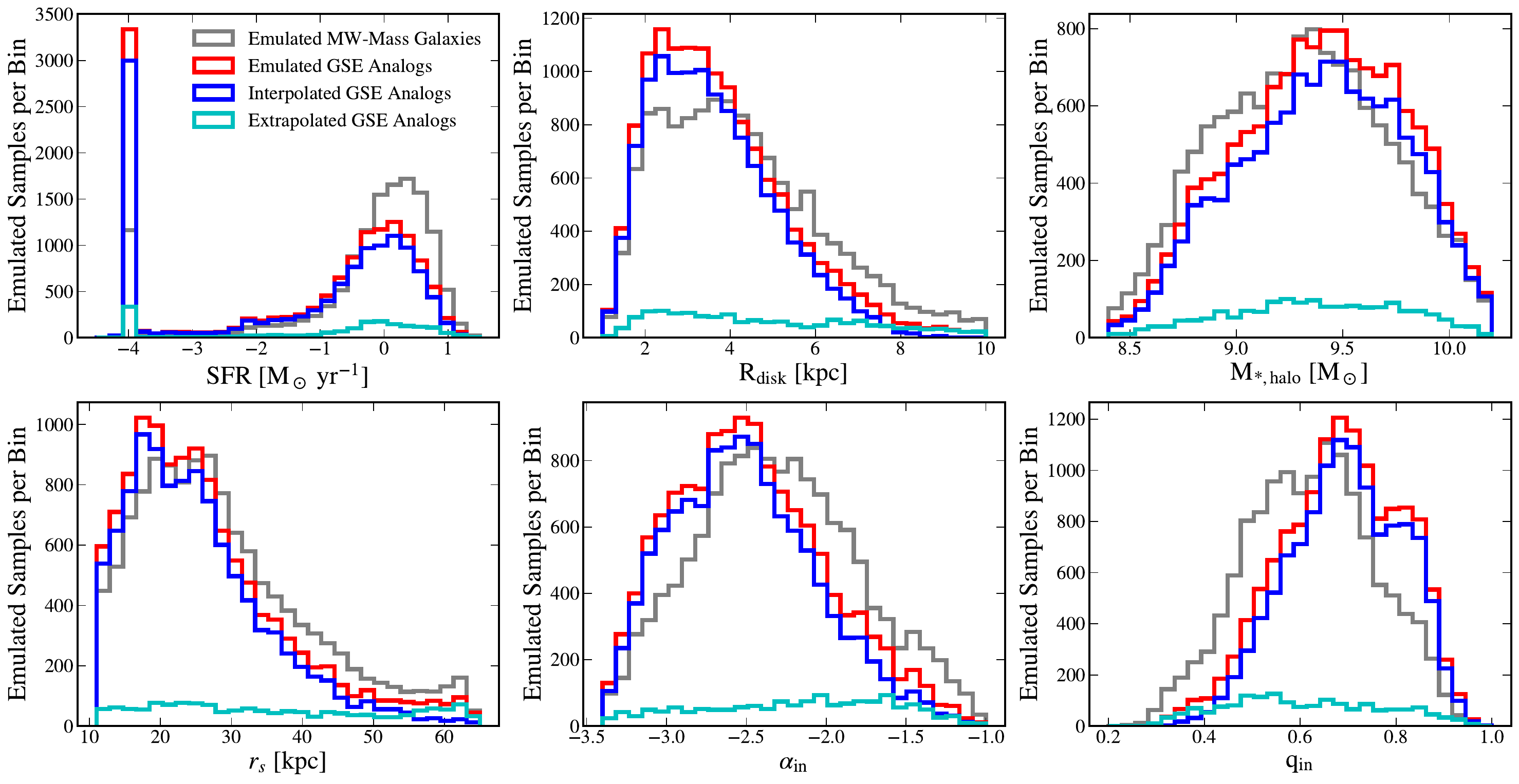}
    \caption{Validation of the extrapolated sample stability.
    The emulated GSE population~(red) is split into interpolated samples~(normalized distance $<2.7$; dark blue) and extrapolated samples~(normalized distance $> 2.7$; cyan) and compared to all emulated MW-mass galaxies~(gray).
    The histograms show where the extrapolated points may exhibit systematically different physical properties (e.g., large values of $r_s$ or small values of $q_{\rm in}$) compared to the interpolated points.
    Generally, the emulator behaves robustly, even at the edges of the learned distribution, but restricting the analysis to only interpolated regions does not change the main results presented in Section~\ref{sec:gse}.
    }
    \label{fig:extrapolation}
\end{figure*}

A critical concern for machine learning models in high-dimensional spaces is the risk of extrapolation, where the emulator samples regions of the latent space that are poorly constrained by the training data.
To quantify this, we analyze the distances between samples in the normalized 13-dimensional space relevant to the GSE analysis.
These parameters are the energy of SN feedback, $\widebar{e}_w$, the speed of SN winds, $\kappa_w$, the scale factor of the last major merger, $t_{\rm mm}$, the scale factor of the candidate GSE merger, $t_{\rm gse}$, the stellar-mass ratio of the MW and candidate GSE merger, $M_{\rm MW}/M_{\rm gse}$, the fraction of \emph{ex-situ} halo stars at $z=0$ from the GSE candidate, $f$, the anisotropy of that debris, $\beta$, the instantaneous star formation rate at $z=0$, SFR, the scale length of the disk,  $R_{\rm disk}$, the stellar halo mass, $M_{*,\mathrm{halo}}$, the stellar halo break radius, $r_s$, the slope of the stellar halo inner-density profile, $\alpha_{\rm in}$, and the shape of the inner stellar halo, $q_{\rm in}$.
Each parameter is normalized such that it has a mean of 0 and a standard deviation of 1 to ensure that one property does not dominate the distance measurement.

Figure~\ref{fig:cdf} shows the cumulative distribution function~(CDF) of the distances to the nearest neighbor in this high-dimensional latent space.
The green line represents the distances between each simulated disk galaxy and the nearest simulated disk galaxy in the DREAMS suite.  This indicates the general distance between any two simulated points in the dataset.
The gold line represents the distances between each of the 11 simulated GSE analogs and the nearest other simulated GSE analog, representing the space of GSE analogs that are sampled by the simulations.
The red line shows the distance between each of the emulated GSE analogs to the nearest of the 11 simulated GSE analogs. 
The overlap between these distributions confirms that the majority of the generated samples lie within the training data distribution, meaning the emulator is effectively interpolating rather than extrapolating (larger distances) or overfitting (smaller distances).

To test the stability of the few samples that lie at the edge of the latent space, Figure~\ref{fig:extrapolation} explicitly separates the emulated population of GSE analogs~(red) into interpolated (normalized distance < 2.7; dark blue) and extrapolated (normalized distance > 2.7; cyan) subsets to compare with all emulated MW-mass galaxies~(gray).
Generally, extrapolated samples do not exhibit unphysical artifacts or discontinuities, although in some cases they do contribute to the tails of the distributions (high $R_{\rm disk}$, high $r_s$, high $\alpha_{\rm in}$, and low $q_{\rm in}$).
However, just because these samples are outside the latent space sampled by the 11 GSE analogs does not mean that they are necessarily hallucinations of the emulator.
The small sample of simulated GSEs in the DREAMS suite means that some parts of the latent space may not be sampled by the simulations, but will be sampled by the emulator.
Additional GSE-analog simulations are needed to ensure that these cases correctly capture the edges of the latent space, which the emulator is extrapolating to based on trends learned from the complete sample.

In summary, our comprehensive validation of the emulators used in this paper provides a faithful representation of the DREAMS simulation data.
The models accurately reproduce the marginalized one-dimensional distributions and the complex two-dimensional covariances of the galaxy populations, capture the underlying physical response of the galaxies to changes in the feedback parameters, and converge with respect to training and emulated sample sizes.
Additionally, our latent space analysis confirms that the emulation process is dominated by interpolation within the supported training data, even when selecting a small sample of 11 simulated GSE analogs.
These tests provide high confidence that the statistical inferences drawn from the emulated galaxy samples in the main text are physically robust and not artifacts of the machine learning method.

\section{Additional Properties for Galaxies with GSE Analogs}
\label{app:props}

\setcounter{equation}{0}
\setcounter{figure}{0} 
\setcounter{table}{0}
\renewcommand{\theequation}{B\arabic{equation}}
\renewcommand{\thefigure}{B\arabic{figure}}
\renewcommand{\thetable}{B\arabic{table}}

Section~\ref{sec:gse} utilizes the emulator ensemble to generate a statistical sample of $10^6$ MW-mass galaxies, allowing us to identify robust sub-populations with a GSE-like event.
This appendix provides a more comprehensive analysis of the structural and kinematic properties of these analogs compared to the general MW-mass galaxy population.  Appendix~\ref{app:gse_props_app} presents supplementary figures showing how the requirement of a GSE-like event affects the properties of the host.  Appendix~\ref{app:TNG} discusses how to disentangle the effects of halo-to-halo variance from uncertainties on the baryonic physics modeling.

\subsection{Detailed Property Distributions}
\label{app:gse_props_app}

\begin{figure*}[b]
    \centering
    \includegraphics[width=\textwidth]{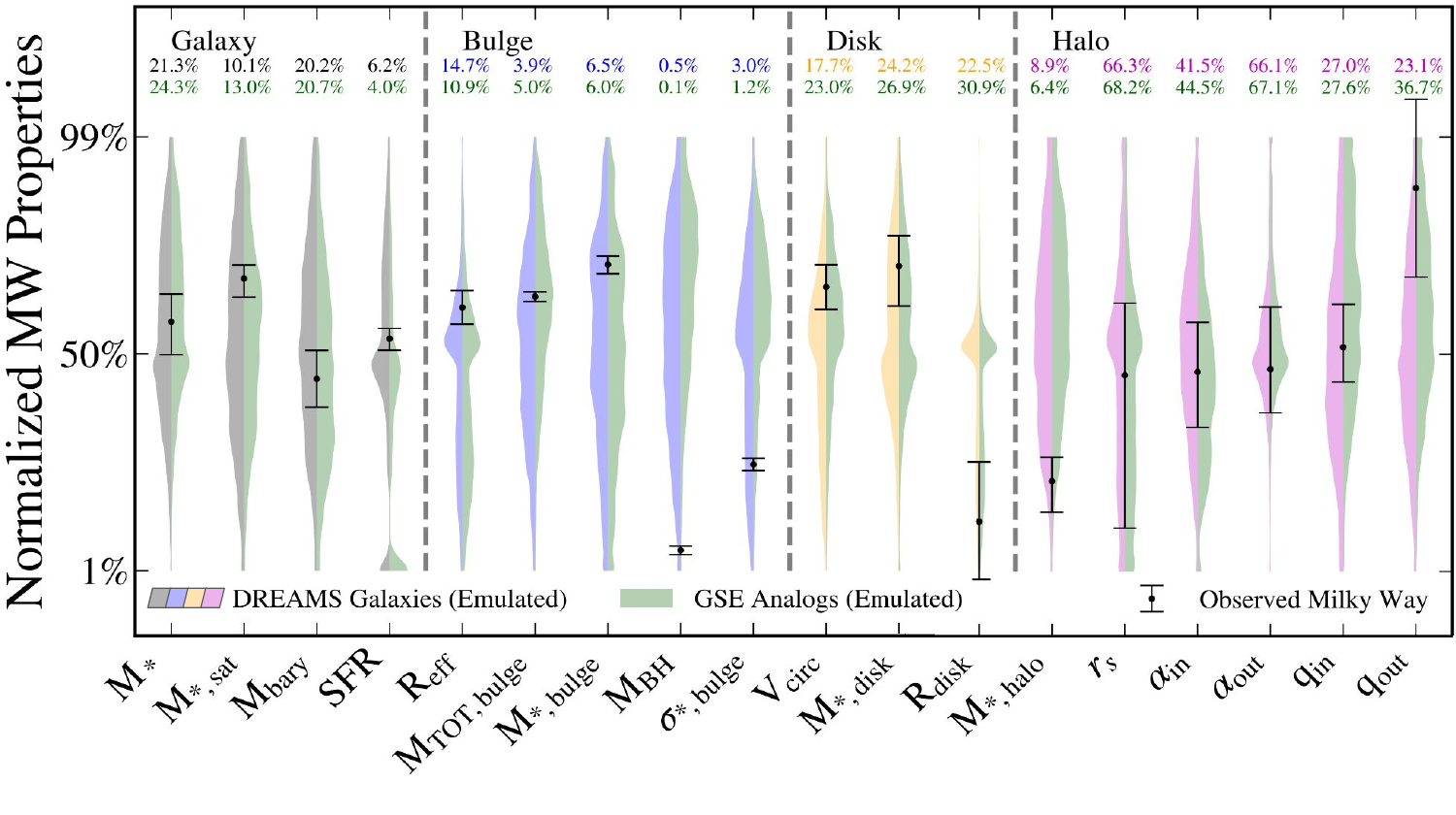}
    \caption{A direct comparison of galaxy properties between the emulated full DREAMS sample (left side of violins) and an emulated subset of galaxies that have a GSE analog (right side of violins).
    Black points show the observed values for the MW taken from~\cite{2016Bland}.
    Text above each violin shows the percentage of each population that falls within 1$\sigma$ of the observed MW value for each property.
    This comparison highlights the significant scatter that remains from halo-to-halo variance.}
    \label{fig:violin_gse}
\end{figure*}

\begin{figure*}[p]
    \centering
    \includegraphics[width=\textwidth]{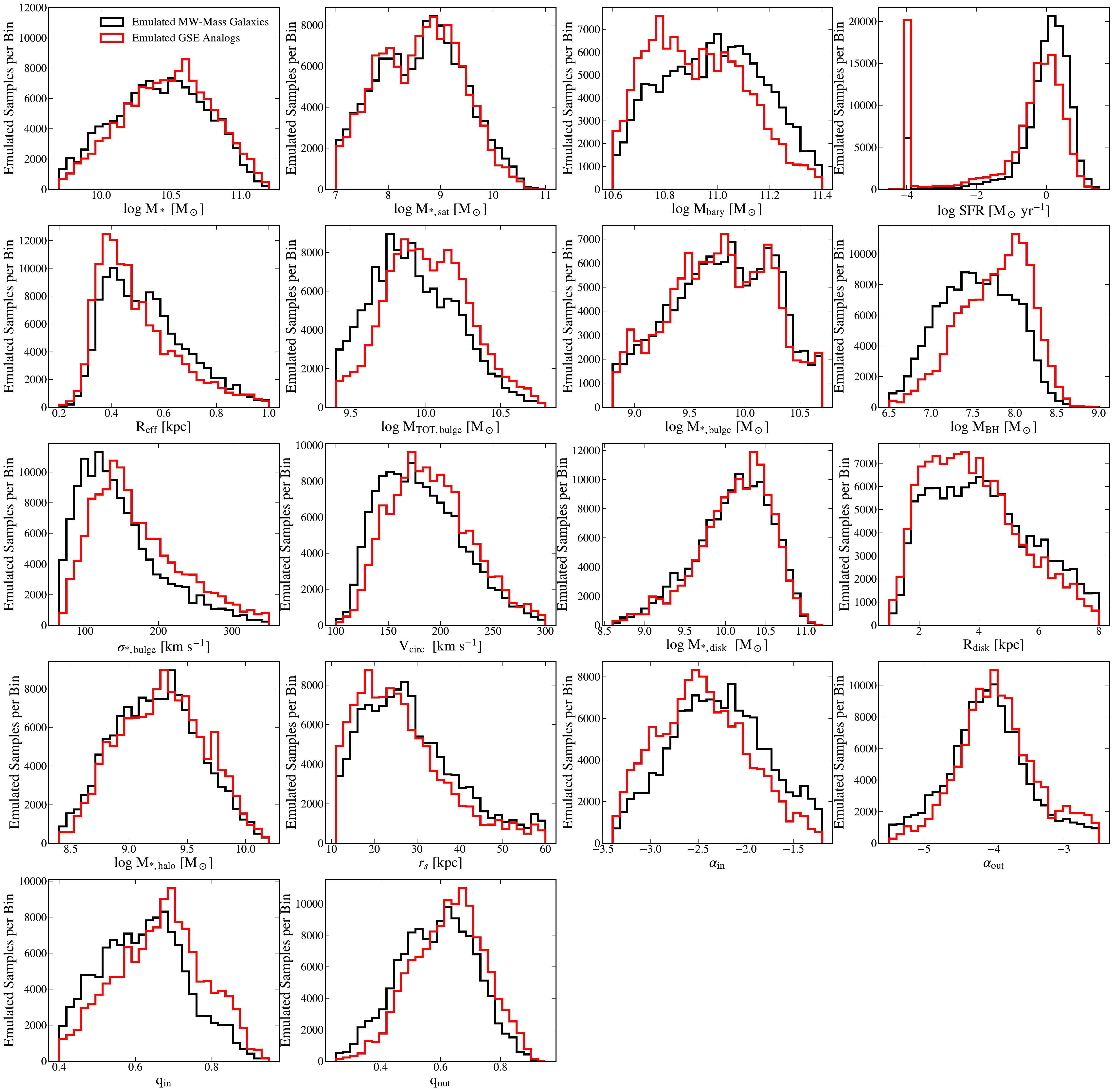}
    \caption{Distributions for all 18 MW properties analyzed in this work for the emulated population of MW-mass galaxies~(black) and the emulated population of galaxies with a GSE analog~(red).
    This overview shows additional properties that are not included in Figure~\ref{fig:gse_props} due to either significant baryon physics dependencies~(such as bulge properties) or little deviation~($M_*$, $M_{*,\mathrm{sat}}$, $M_{*,\mathrm{disk}}$, $\alpha_{\rm out}$).
    }
    \label{fig:1D_props}
\end{figure*}

Figure~\ref{fig:violin_gse} provides an analogous plot to the one shown in Figure~\ref{fig:violin}.
The population of the emulated MW-mass galaxies (colored violins on left) is compared to the population of emulated galaxies with a GSE analog (green violins on right), with the percentage of each that fall within the 1$\sigma$ observational range on top.
We note that the left violins differ slightly from those shown in Figure~\ref{fig:violin} because they now correspond to an emulated sample instead of the DREAMS simulations.
The comparison reveals that selecting for a GSE-like merger history can improve the agreement with the MW for specific structural properties.
The most notable improvements are in the outer halo shape ($q_{\rm out}$) where the overlap increases from 23.1\% to 36.7\% and the disk scale length ($R_{\rm disk}$) where the overlap increases from 19\% to 25\%.
These shifts support the hypothesis that the MW's specific assembly history plays a role in shaping its current morphology.

Figure~\ref{fig:1D_props} presents the probability density functions for the 18 galactic properties analyzed in this work (see Table~\ref{tab:props}). The black histograms represent the complete, weighted population of DREAMS galaxies, while the red histograms represent the subset that satisfies the four GSE selection criteria ($0.33<t_{\rm gse}<0.5$, $t_{\rm mm}<t_{\rm gse}$, $M_{\rm MW}/M_{\rm gse}<5$, $f<0.5$, and $\beta<0.5$).
Apart from the shifts to the stellar disk and inner stellar halo discussed in the main text, other notable shifts include lower baryon masses~($M_{\rm bary}$), smaller bulge effective radii~($R_{\rm eff}$), larger bulge total mass~($M_{\rm TOT, bulge}$), more massive SMBHs~($M_{\rm BH}$), greater stellar bulge velocity dispersions~($\sigma_{*, \mathrm{bulge}}$), higher circular velocities~($V_{\rm circ}$), and a more spherical inner/outer halo~($q_{\rm in}$, $q_{\rm out}$).
In all these cases, the size of the shifts are not dramatic, given the overall dispersion of the distributions.  In some cases (such as $M_{\rm BH}$), these trends are not directly linked to the GSE merger and instead are a consequence of its dependence on \ew; see Appendix~\ref{app:TNG} for more details.

In contrast, other properties show little difference.
As noted in the main text, the most notable of these is the stellar halo mass ($M_{\rm *, halo}$), which does not exhibit a systematic enhancement for the GSE population.
Other properties that do not shift between these populations include the total stellar mass~($M_*$), the stellar mass in satellites~($M_{\mathrm{*,sat}}$), the stellar mass of the bulge~($M_\mathrm{*,bulge}$), the stellar mass of the disk~($M_{\rm *,disk}$), and the slope of the surface-density profile for the outer stellar halo~($\alpha_{\rm out}$).
These results suggest that, especially for the stellar mass of the bulge, disk, and halo, a more detailed match to the MW's merger history and \emph{in-situ} star formation history is needed to match the stellar mass content of the MW.
Additionally, the substantial scatter remaining in all properties of the GSE subset underscores the dominant role of stochastic halo-to-halo variance and the need for a more specific definition of a MW-analog that includes selections beyond present-day mass and the GSE merger.

\begin{figure*}[p]
    \centering
    \includegraphics[width=\textwidth]{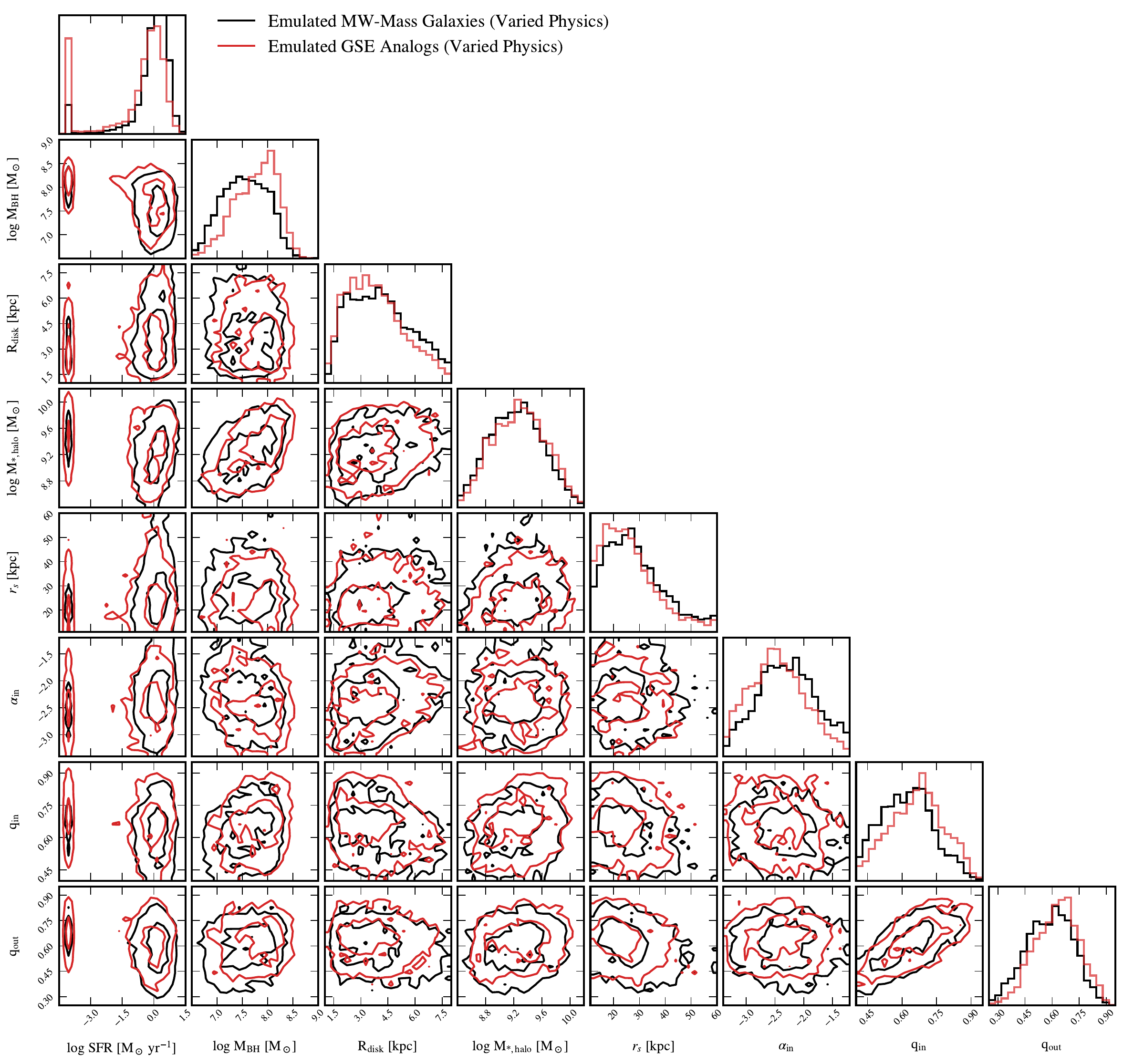}
    \caption{This plot compares the unnormalized distributions of key structural properties for the emulated MW-mass galaxy population~(black) with the rare subset of GSE analogs~(red).
    It provides the underlying data for Figure~\ref{fig:violin_gse} and demonstrates the shift in distributions between the general population and the subset with a GSE analog.  The two-dimensional contours correspond to 1 and 2$\sigma$ containment regions.
    }
    \label{fig:nehod_corner_gse}
\end{figure*}

Figure~\ref{fig:nehod_corner_gse} displays the covariance between key properties of the galaxies with a GSE analog.
The contours highlight that while the GSE analogs~(red) occupy a distinct region of the high-dimensional galactic property space, they do not form a disjoint island.
Instead, they represent a biased subset of the general population~(black).
For example, in the SFR-$q_{\rm in}$ plane, the GSE analogs are concentrated in the region of low star formation and spherical stellar halos, yet they overlap significantly with the low-SFR tail of the general distribution.
This overlap emphasizes that while a GSE merger creates favorable conditions for these properties, which could be through gas depletion or angular momentum redistribution, alternative evolution pathways can produce similar $z=0$ states.

\subsection{Isolating the Effects of Baryonic Feedback}
\label{app:TNG}

\begin{figure*}
    \centering
\includegraphics[width=\textwidth]{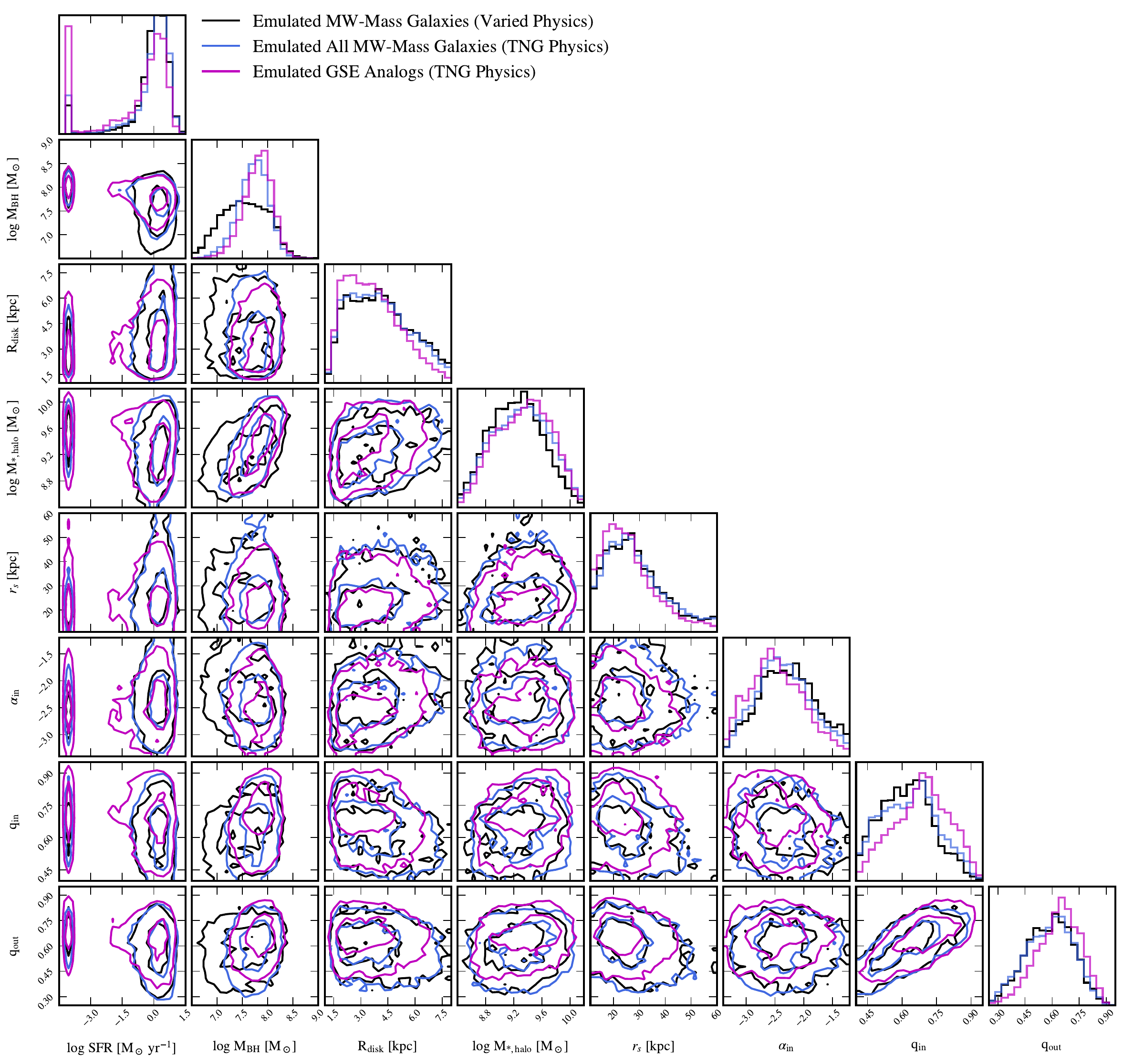}
\caption{This figure demonstrates the emulator's ability to isolate effects of different sources of variation on galactic properties.
It compares the distributions for three distinct datasets: all emulated galaxies with varied physics~(black), all emulated galaxies with fixed TNG physics~(blue), and emulated galaxies with a GSE analog and fixed baryonic physics~(purple).
This comparison allows one to separate impacts from model uncertainty and the GSE merger.}
\label{fig:nehod_corner_tng}
\end{figure*}

A central challenge in interpreting the properties of the MW is determining whether its features arise from its specific assembly history (like the GSE merger) or from the specific baryonic feedback processes implemented.
The emulator we employ offers a unique opportunity to break this degeneracy by performing a `fixed physics' test.
We generate a new population of $10^6$ samples where the astrophysical parameters are held constant at their fiducial TNG values \citep[$\Omega_{\rm m}=0.314$, $\sigma_8=0.834$, $\widebar{e}_w=3.6$, $\kappa_w=7.4$, and $\epsilon_{f,\mathrm{high}}=0.1$;][]{2018Pillepicha}, varying only the total mass of the MW~($\mtot$).

Figure~\ref{fig:nehod_corner_tng} isolates these effects by comparing three distributions: the general population with varied physics~(black), the general population with fixed TNG physics (blue), and the population with a GSE analog and fixed TNG physics~(purple).
This comparison reveals two distinct classes of behavior.
First is the case where property shifts are driven by the baryonic physics assumptions, rather than the merger history.  In these cases, the blue histograms look different from the black, but closely align with the purple.   
This is most evident for the bulge components, such as $M_{\rm BH}$.
Additional properties where this is observed (not shown for readability) are $M_{\rm bary}$, $R_{\rm eff}$, $M_{\rm TOT,bulge}$, $\sigma_{\rm *,bulge}$, and $V_{\rm circ}$.
Likely, the apparent shifts in the GSE population shown in Figure~\ref{fig:1D_props} for these properties are due to selection effects: selecting GSE analogs preferentially biases one towards lower SN energies (see Figure~\ref{fig:gse_prob}), which can then impact properties like the BH mass. 

For the second case, fixing to the TNG physics does not change the property distribution (blue and black histograms are similar), but requiring a GSE analog does (blue and purple histograms differ).  
Examples include the SFR, the disk scale length~($R_{\rm disk}$), and stellar halo shape parameters~($q_{\rm in}$ and $q_{\rm out}$).
By isolating these history-driven features from feedback-driven scatter, the emulator provides a rigorous physical basis for identifying true signatures of the MW's formation history.

\clearpage
\bibliography{citations}
\bibliographystyle{aasjournal}

\label{lastpage}
\end{document}